\documentclass[11pt,a4paper]{article}
\pdfoutput=1
\usepackage{mystyle}
\usepackage{afterpage}
\usepackage{xcolor}
\usepackage{epsfig}
\usepackage{cancel}
\usepackage{framed}
\usepackage{multirow}

\usepackage{seqsplit}
\usepackage{enumitem} 

\usepackage{amsmath, amsthm, amsfonts, amssymb, latexsym}
\usepackage{upgreek}
\usepackage[caption=false]{subfig}

\usepackage{url}
\usepackage{ifpdf}

\def\logbar{\overline{\log\,}}

\def\beq{\begin{equation}}
\def\eeq{\end{equation}}
\def\bequ{\begin{equation}}
\def\eequ{\end{equation}}
\title{NLO electroweak corrections in general scalar singlet models}

\author[a,b]{Raul Costa,}
\emailAdd{raul.costa@cern.ch}
\author[b]{Marco O. P. Sampaio,} 
\emailAdd{msampaio@ua.pt}
\author[c,d,e]{Rui Santos}
\emailAdd{rasantos@fc.ul.pt}

\affiliation[a]{CERN, Geneva, Switzerland} 
\affiliation[b]{Departamento de F\'\i sica da Universidade de Aveiro and CIDMA (Center for Research $\&$ Development in Mathematics and Applications) \\ 
Campus de Santiago, 3810-183 Aveiro, Portugal} 
\affiliation[c]{
 Instituto Superior de Engenharia de Lisboa - ISEL \\
 1959-007 Lisboa, Portugal
}
\affiliation[d]{Centro de F\'\i sica Te\'orica e Computacional,
Universidade de Lisboa  \\
1649-003 Lisboa, Portugal} 
\affiliation[e]{
 LIP, Departamento de F\'{\i}sica, Universidade do Minho, \\
 4710-057 Braga, Portugal}

\keywords{Higgs Physics, Beyond Standard Model, Renormalisation}

\abstract{If no new physics signals are found, in the coming years, at the Large Hadron Collider Run-2, an increase in precision of the Higgs couplings measurements will shift the dicussion to the effects of higher
order corrections. In Beyond the Standard Model (BSM) theories this may become the only tool to probe new physics. Extensions of the Standard Model (SM) with several scalar singlets may address several of its problems, namely to explain dark matter, the matter-antimatter asymmetry, or to improve the stability of the SM up to the Planck scale. In this work we propose a general framework to calculate one loop-corrections in BSM models with an arbitrary number of scalar singlets. We then apply our method to a real and to a complex scalar singlet models. We assess the importance of the one-loop radiative corrections first by computing them for a tree level mixing sum constraint, and then for the main Higgs production process $gg \to H$. We conclude that, for the currently allowed parameter space of these models, the corrections can be at most a few percent. Notably, a non-zero correction can survive when dark matter is present, in the SM-like limit of the Higgs couplings to other SM particles.
}
\begin{document}

\maketitle

\section{Introduction}
\label{sec:Introduction}

With the start of Run 2, CERN's Large Hadron Collider (LHC) has entered the stage of precision measurements
of the Higgs couplings to the Standard Model (SM) particles. Even though the particle physics community is focused
on the search for direct signals of beyond the SM (BSM) physics, it may happen that no such signal 
is detected during Run 2. If this is the case, we need to take advantage of the precise determination of 
the relevant Higgs couplings to understand if any new physics contributions can be hidden behind those measurements. 
Scalar extensions of the SM have, in most cases, a decoupling limit where, if the new scalar states are heavy enough, 
the model can only be probed via radiative corrections. In fact, if we are faced with a situation where no direct
hint of new physics is found, manifestations of BSM physics can only appear through deviations in the measured Higgs couplings.

In this work we will focus on extensions of the SM where an arbitrary number of singlets is added to the
SM field content. These are the simplest extensions of the scalar sector that introduce a dark matter 
candidate~\cite{Silveira:1985rk, McDonald:1993ex, Burgess:2000yq, Bento:2000ah, Davoudiasl:2004be, 
Kusenko:2006rh, vanderBij:2006ne, He:2008qm, Gonderinger:2009jp, Mambrini:2011ik, He:2011gc, Gonderinger:2012rd, Cline:2013gha, Gabrielli:2013hma, Profumo:2014opa}. These models also allow for a strong first-order phase transition during the era of 
Electroweak Symmetry Breaking (EWSB) if
the extension comprises at least two singlets~\cite{Menon:2004wv, Huber:2006wf, Profumo:2007wc,Barger:2011vm, Espinosa:2011ax}.
Hence, at least two of the outstanding problems of the SM can be solved within the framework of these models, namely a candidate for dark matter and a solution to the matter antimatter asymmetry via electroweak baryogenesis. It should be noted that extensions with only one dark scalar singlet are basically excluded by the latest LUX results~\cite{Akerib:2016vxi} when combined with the requirement that the dark matter relic density of the model matches the one obtained from the Cosmic Microwave background data. We have verified that this is not the case when at least two new singlets are added to the SM, one of them being dark and the other mixing with the SM-like Higgs.

If the LHC indeed does not find strong signs of new physics, such as new particle states, the scale for such new physics may be as large as the GUT or the Planck scales. This energy is unattainable by any current or planned collider experiments so we may have to work in a framework that is a good description of the fundamental interactions up to some high energy scale. Thus, any effective description that improves theoretical problems of the SM is an interesting candidate.  In a previous work we have shown that the complex singlet extension of the SM also improves the stability of the SM. In fact, the presence of a heavier 
scalar state, which has to be heavier than about $140$~GeV, can stabilise the SM up to the Planck scale~\cite{Costa:2014qga}.  

In this article we focus on the issue of determining electroweak (EW) radiative corrections in general scalar SM extensions, with emphasis on the scalar singlet models framework. Our main goal is to find a general set of expressions that allows us to obtain next to leading order (NLO) electroweak corrections to the parameters of a model with any number of scalar singlet fields. We will go beyond the effective potential approach recently studied in~\cite{Camargo-Molina:2016moz}, which is valid only when the new degrees of freedom are heavy. Thus, though we formulate our results to connect to that limit, they are valid for any external momentum scale (contrarily to the effective potential approximation which is valid for small external momenta). In our framework we obtain a set of conditions consistently truncated in an expansion in powers of $\hbar$ which, once a number of consistent independent input parameters are chosen, deliver the NLO EW corrections for the remaining parameters. A special attention is given to the treatment of tadpoles and propagators and we provide a generic strategy to easily transform between different schemes. In connection with the effective potential approximation we also discuss, on general grounds, the issue of infrared divergences. We then apply our method to the real scalar singlet extension (RxSM) and to the complex scalar singlet extension (CxSM) of the SM. However, we note that the method is ready to be applied to SM extensions with an arbitrary number of singlet fields and that many of our formulas are also useful for other scalar extensions of the SM. In particular, our approach is especially suited for the automation of the computation of higher order corrections in general purpose numerical tools to scan the parameter space of scalar extensions of the SM~\cite{ScannerS,Coimbra:2013qq}.

Higher order corrections to singlet extensions of the SM have been performed in~\cite{Kanemura:2015fra, Bojarski:2015kra}.
The corrections to the SM-like Higgs coupling to fermions and gauge bosons was shown to be of the order of  1 \%~~\cite{Kanemura:2015fra}. Furthermore the corrections were maximal in the decoupling limit where the model becomes indistinguishable from the SM. 
Electroweak corrections to the decay $H \to hh$ were performed in ~\cite{Bojarski:2015kra}. With the main theoretical and 
experimental constraints taken into account, it was shown that corrections to the triple scalar vertex ($Hhh$)
are of the order of a few percent. Calculations of higher order corrections in the complex singlet
extension of the SM are still not available. With this work we will not only present a set of equations 
to renormalise the parameters of the theory at one loop but we will then also use them to calculate the electroweak correction to Higgs production via gluon fusion. This last calculation is performed near the decoupling limit with the main purpose to understand the contributions of the triple scalar couplings of the various scalars running in the loops at NLO. Clearly, with all the SM-like Higgs coupling close
to the SM ones, the only large effects in the radiative corrections would have to come from such scalar-scalar interactions. The numerical
analysis in our examples will be performed for three particular cases: the broken RxSM, with a new Higgs boson mixing with the SM-like one, and the broken and symmetric CxSM with, respectively, three mixing Higgs bosons, and two mixing Higgs bosons and a dark matter scalar. We will find that, consistently with earlier calculations for the NLO corrections to the decays, the corrections are very small, of the order of a few percent, also for production. Nevertheless, we will find that the presence of a dark matter particle can enhance the corrections, even very close to the SM-like limit, compared with the other models (though still in the few percent order).

The smallness of the electroweak corrections in the real singlet models calls for prudence in the claims
of measurable differences relative to SM Higgs couplings. The interference effects for this kind of BSM 
scenarios was first addressed in~\cite{Maina:2015ela} for the real singlet model, showing that
interference effects to $gg \to h^*, H^{(*)} \to ZZ \to 4l$ can be important away from the non-SM
scalar ($H$) peak region. Although the interference effects can be of up to order ${\cal O}$(1)
for the integrated cross sections for the 8 TeV LHC~\cite{Kauer:2015hia}, judicious kinematical cuts
can reduce the interference effects to ${\cal O}$(10\%). 
Interference effects at NLO QCD were discussed in~\cite{Dawson:2015haa}
for the process $gg \to h^*, H^{(*)} \to hh$. It was shown that the
double Higgs invariant mass can increase by up to 20\% or decrease 
by up to 30\% depending of the heavier Higgs mass. More importantly,
interference effects can significantly distort the kinematic distribution
around the resonant peak of the heavy Higgs. Recently the effects
of higher order operators in the real singlet model~\cite{Dawson:2016ugw}
again showed that large cancellations can occur due to interference
effects between the two sectors. In conclusion, if a significant deviation
is found in Higgs couplings, the radiative corrections have to be combined
with the interference for a proper interpretation of the results.  

The structure of the paper is as follows. In the first two sections we start by defining our strategy. We present the Lagrangians and fix the notation  in Sect.~\ref{sec:DefsNotation} and then, in Sect.~\ref{sec:conventions},
we obtain our master linear system that, given a choice of input parameters, provides as output the remaining renormalised parameters at NLO EW. The issue of infrared divergences in connection with the effective potential approximation  is discussed in Sects.~\ref{subsec:Coleman-Weinberg} and~\ref{subsec:IRbehaviour}. In Sect.~\ref{sec:applications} we apply the procedure first to a general class of scalar singlet extensions of the SM and then specialise to the RxSM and to the CxSM, Sect.~\ref{subsec:particular_models}, for which we provide a numerical analysis in Sect~\ref{sec:Results}. Our conclusions are summed up in Sect.~\ref{sec:conclusions} and several useful formulae/derivations are provided in the appendices.

\section{Definitions and notation}
\label{sec:DefsNotation}

To define a general four dimensional gauged Quantum Field Theory (QFT) Lagrangian we use the notation of~\cite{Martin:2001vx} with a few adaptations~\cite{Camargo-Molina:2016moz,Costa:2014qga}. We assume a decomposition of a general renormalisable Lagrangian where the gauge basis fields are such that: i) all scalar field multiplets are decomposed as $N_0$ canonically normalised real scalar fields, $\Phi_i$ ($i=1,\ldots,N_0$), ii) all fermion multiplets are decomposed as a set of $N_{1/2}$ two-component Weyl fermions, $\Psi_I$ ($I=1,\ldots,N_{1/2}$) and iii) there are $N_1$ gauge bosons in the adjoint representation of the gauge group, i.e. $\mathcal{A}_a^{\mu}$ ($a=1,\ldots,N_1$). We adopt the Einstein convention where repeated indices which are one up (superscript) and one down (subscript) are summed over. If the repeated indices are both down or both up they are \underline{not} summed over.  All (non-spacetime) latin indices are assumed to be in Euclidean space -- they are lowered and raised with the identity matrix. The gauge basis interaction Lagrangian (i.e. suppressing kinetic terms) is then composed of the following terms:
\begin{eqnarray}
-\mathcal{L}_{S} & = & L^{i}\Phi_{i}+\dfrac{1}{2}L^{ij}\Phi_{i}\Phi_{j}+\dfrac{1}{3!}L^{ijk}\Phi_{i}\Phi_{j}\Phi_{k}+\dfrac{1}{4!}L^{ijkl}\Phi_{i}\Phi_{j}\Phi_{k}\Phi_{l}\nonumber\\
-\mathcal{L}_{F} & = & \frac{1}{2}Y^{IJ}\Psi_{I}\Psi_{J}+\frac{1}{2}Y^{IJk}\Psi_{I}\Psi_{J}\Phi_{k}+c.c. \label{Eq:L-basis}\\
-\mathcal{L}_{SG} & = & \dfrac{1}{4}G^{abij}\mathcal{A}_{a\mu}\mathcal{A}_{b}^{\mu}\Phi_{i}\Phi_{j}+G^{aij}\mathcal{A}_{a\mu}\Phi_{i}\partial^{\mu}\Phi_{j}\nonumber\\
-\mathcal{L}_{FG}	&=&	-G_{{\phantom{a}}I}^{a\phantom{I}J}\mathcal{A}_{a\mu}\Psi^{\dagger I}\bar{\sigma}^{\mu}\Psi_{J} \nonumber\\
-\mathcal{L}_{G} & = & -G^{abc}\mathcal{A}_{a\mu}\mathcal{A}_{b\nu}\partial^{\mu}\mathcal{A}_{c}^{\nu}+\dfrac{1}{4}G^{abe}G_{\phantom{cd}e}^{cd}\mathcal{A}_{a}^{\mu}\mathcal{A}_{b}^{\nu}\mathcal{A}_{c\mu}\mathcal{A}_{d\nu}-G^{abc}\mathcal{A}_{a\mu}\omega_{b}\partial^{\mu}\bar{\omega}_{c} \; ,\nonumber
\end{eqnarray}
where the ghost fields are represented by $\omega_a$ and $c.c.$ denotes complex conjugation. We call this the $L$-basis following the nomenclature in~\cite{Camargo-Molina:2016moz,Costa:2014qga}, where the pure scalar, fermionic and gauge interaction coupling tensors are denoted, respectively by $\{L^{\ldots},Y^{\ldots},G^{\ldots}\}$  with $\ldots$ replaced by suitable sets of indices. Note that for a simple gauge group $G^{abc}$  is given by $G^{abc}=g\,f^{abc}$ with $g$ the gauge coupling constant and $f^{abc}$ the structure constants of the gauge group. For a direct product group we can still encode all information in $G^{abc}$ by requiring a block structure. This can be represented using sub-ranges for the indices $a_1=1,\ldots,n_1$, $a_2=n_1+1,\ldots,n_3$, etc... if components that have indices not all in the same sub-range are zero. More concretely we would have $G^{a_1b_1c_1}=g_1f_1^{a_1b_1c_1}$, $G^{a_2b_2c_2}=g_2f_2^{a_2b_2c_2}$, etc... and, for example, $G^{a_1b_2c_2}=0$.

A second form of the interaction terms is obtained after shifting the scalar fields by a general classical field configuration such that $\Phi_i(x)=v_i+\phi_i(x)$:
\begin{eqnarray}
-\mathcal{L}_{S} & = & \Lambda+\Lambda^{i}_{(S)}\phi_{i}+\dfrac{1}{2}\Lambda^{ij}_{(S)}\phi_{i}\phi_{j}+\dfrac{1}{3!}\Lambda^{ijk}_{(S)}\phi_{i}\phi_{j}\phi_{k}+\dfrac{1}{4!}\Lambda^{ijkl}_{(S)}\phi_{i}\phi_{j}\phi_{k}\phi_{l}\nonumber\\
-\mathcal{L}_{F} & = & \frac{1}{2}M^{IJ}\Psi_{I}\Psi_{J}+\frac{1}{2}Y^{IJk}\Psi_{I}\Psi_{J}\phi_{k}+c.c \label{Eq.Lambda-basis}\\
-\mathcal{L}_{SG} & = & \dfrac{1}{2}\Lambda^{ab}_{(G)}\mathcal{A}_{a\mu}\mathcal{A}_{b}^{\mu}+\dfrac{1}{2}\Lambda^{abi}_{(G)}\mathcal{A}_{a\mu}\mathcal{A}_{b}^{\mu}\phi_{i}+\dfrac{1}{4}\Lambda^{abij}_{(G)}\mathcal{A}_{a\mu}\mathcal{A}_{b}^{\mu}\phi_{i}\phi_{j}+G^{aij}\mathcal{A}_{a\mu}\phi_{i}\partial^{\mu}\phi_{j}\nonumber\\
-\mathcal{L}_{FG}	&=&	-G_{{\phantom{a}}I}^{a\phantom{I}J}\mathcal{A}_{a\mu}\Psi^{\dagger I}\bar{\sigma}^{\mu}\Psi_{J} \nonumber\\
-\mathcal{L}_{G} & = & -G^{abc}\mathcal{A}_{a\mu}\mathcal{A}_{b\nu}\partial^{\mu}\mathcal{A}_{c}^{\nu}+\dfrac{1}{4}G^{abe}G_{\phantom{cd}e}^{cd}\mathcal{A}_{a}^{\mu}\mathcal{A}_{b}^{\nu}\mathcal{A}_{c\mu}\mathcal{A}_{d\nu}-G^{abc}\mathcal{A}_{a\mu}\omega_{b}\partial^{\mu}\bar{\omega}_{c}\; .\nonumber
\end{eqnarray}
Here we have introduced the notation $\Lambda_{(T)}^{\ldots}$ for the interaction coupling tensors containing a field of type $T$ and scalar fields without derivatives where the type $T$ runs over the three possible types of fields $\{S,F,G\}$ (scalar, fermionic and gauge respectively) and $\ldots$ represents a set of indices. These couplings appear in the calculation of the effective potential which, at one loop in the Landau gauge, can be organised as a sum over field types -- to be discussed in Sect.~\ref{subsec:Coleman-Weinberg}. For fermions the natural objects appearing in the  Eq.~\eqref{Eq.Lambda-basis} after the shift are the mass matrix $M^{IJ}$ and the Yukawa couplings $Y^{IJk}$. However, we can also define a mass squared matrix and (effective) cubic and quartic couplings $\{\Lambda^{IJ}_{(F)},\Lambda^{IJk}_{(F)},\Lambda^{IJkm}_{(F)}\}$. The latter can also be found in appendix~\ref{app:RelBases} together with all other $v_i$ dependent parameters, $\Lambda_{(T)}^{\ldots}$ -- see also~\cite{Camargo-Molina:2016moz}.
 Finally, we can rotate all fields to their mass eigen-basis (named the $\lambda$-basis) through orthogonal or unitary transformations, for bosons and fermions respectively. Using the transformations~\eqref{Eq:Rot-fields} then we have 
\begin{eqnarray}
-\mathcal{L}_{S}	&=&	\Lambda+\lambda^{i}_{(S)}R_{i}+\dfrac{1}{2}\left(m^{i}_{(S)}\right)^{2}R_{i}^{2}+\dfrac{1}{3!}\lambda^{ijk}_{(S)}R_{i}R_{j}R_{k}+\dfrac{1}{4!}\lambda^{ijkl}_{(S)}R_{i}R_{j}R_{k}R_{l} \nonumber\\
-\mathcal{L}_{F}	&=&	\frac{1}{2}m^{IJ}\psi_{I}\psi_{J}+\frac{1}{2}y^{IJk}\psi_{I}\psi_{J}R_{k}+c.c \nonumber\\
-\mathcal{L}_{SG}	&=&	\dfrac{1}{2}\left(m^{a}_{(G)}\right)^{2}A_{a\mu}A_{a}^{\mu}+\dfrac{1}{2}\lambda^{abi}_{(G)}A_{a\mu}A_{b}^{\mu}R_{i}+\dfrac{1}{4}\lambda^{abij}_{(G)}A_{a\mu}A_{b}^{\mu}R_{i}R_{j}+g^{aij}A_{a\mu}R_{i}\partial^{\mu}R_{j} \nonumber\\
-\mathcal{L}_{FG}	&=&	-g_{{\phantom{a}}I}^{a\phantom{I}J}A_{a\mu}\psi^{\dagger I}\bar{\sigma}^{\mu}\psi_{J} \nonumber\\
-\mathcal{L}_{G}	&=&	-g^{abc}A_{a\mu}A_{b\nu}\partial^{\mu}A_{c}^{\nu}+\dfrac{1}{4}g^{abe}g_{\phantom{cd}e}^{cd}A_{a}^{\mu}A_{b}^{\nu}A_{c\mu}A_{d\nu}-g^{abc}A_{a\mu}\omega_{b}\partial^{\mu}\bar{\omega}_{c} \, \label{Eq:lambda_basis}.
\end{eqnarray}
 Note that all couplings in the $\lambda$-basis, Eq.~\eqref{Eq:lambda_basis}, whose indices are rotated according to the transformations induced by \eqref{Eq:Rot-fields},  are now in lower case. For completeness, we provide in appendix~\ref{app:RelBases} the relations between the various bases including the rotation matrices to obtain the mass eigenstates. We note that the latter can be represented collectively, for the field type $T=\{S,G,F\}$, by $\mathcal{U}_{(T)}$ (unitary or orthogonal) with the defining relation that all mass matrices are diagonalised:
\begin{equation}\label{eq:UT}
\mathcal{U}_{(T)}\Lambda_{(T)}\mathcal{U}_{(T)}^\dagger={\rm diag}\{m_{(T)a}^2\} \; .
\end{equation}
On the right hand side of Eq.~\eqref{eq:UT} we use latin indices from the beginning of the alphabet to denote the component of the diagonal. Whenever $T$ is not specified we follow this convention, i.e. we use lower case indices from the beginning of the latin alphabet ($a,b,c,\ldots$) and reserve  indices from the middle of the alphabet ($i,j,k,\ldots$) for scalar field indices\footnote{The latin indices from the beginning of the Latin alphabet will also be used later for the gauge field indices. However there is no danger of confusion because whenever we expand the expressions in the field type $T$ all types of indices appear explicitly (scalar, fermionic and gauge indices).}. Note that whenever we use a matrix notation without explicit indices, $\Lambda_{(T)}$ is assumed to represent $\Lambda_{(T)}^{ab}$, i.e. the mass squared matrix.

In this article we will need to compute the one-loop radiative corrections to the scalar mass eigenstates. These are determined by the poles of the radiatively corrected propagator, $G_{ij}$, between scalar states $i$ and $j$. It is well known that the Dyson re-summed inverse propagator $\left[G^{-1}\right]_{ij}$ can be expressed as~\cite{Ellis:1990nz} 
\begin{equation}\label{eq:inverse_propagator}
i\left[G^{-1}\right]_{ij}(p^2)=p^{2}\delta_{ij}-\partial_{ij}^{2}V_{{\rm eff}}-\Delta\Sigma_{ij}\left(p^{2}\right)
\end{equation}
where $p_{\mu}$ is the external 4-momentum\footnote{We are using the metric signature convention $(+ - - -)$.}. Furthermore 
\begin{equation}
\Sigma_{ij}\left(p^{2}\right)=\delta_{ij}T_{j}\left(p^{2}\right)+\Pi_{ij}\left(p^{2}\right)
\end{equation}
 where $T_{j}$ and $\Pi_{ij}$, are the one-particle irreducible
tadpole (1-point) and self-energy (2-point) functions. The tadpole term can in practice be set to zero by assuming an expansion of the theory around a minimum of the effective potential order by order in perturbation theory~\cite{Martin:2003it} if one works in Landau gauge (which we assume in this work). Finally we have defined $\Delta \Sigma_{ij}\left(p^2\right)\equiv \Sigma_{ij}\left(p^2\right)-\Sigma_{ij}(0)$. The physical pole state $x$ (labelling the $N_0$ physical scalar states) is defined to have an eigenvalue $p^{2}=M_{x}^{2}$ and an eigenvector $E_{\phantom{i}x}^{i}$ such that the pole conditions are obeyed: 
\begin{equation}\label{eq:GenPoleConditions}
\left[G^{-1}\right]_{ij}\left(M_{x}^{2}\right)E_{\phantom{i}x}^{i}=0\:\Rightarrow\det\left[M_x^{2}\mathbf{1}-{\bf \partial^{2}}V_{{\rm eff}}-\Delta{\bf \Sigma}\left(M_x^{2}\right)\right]=0\:.
\end{equation}
Here, we have suppressed the scalar indices inside the determinant
so a matrix notation is used -- see Eq.~\eqref{eq:inverse_propagator}. In general, for unstable particles, the
eigenvalue can have an imaginary part which relates to the width of
the particle so one defines 
\[
M_{x}^{2}\equiv m_{x}^{2}-i\Gamma_{x}m_{x}
\]
where $m_{x}$ is the physical mass and $\Gamma_{x}$ is the width of the particle.

Finally, we will also need to extract  the wave function renormalisation factor, $Z_x$, from the inverse propagator. Projecting the inverse propagator along the normalised eigen-vector this is obtained from
\begin{equation}
G^{-1}_x\equiv \left[G^{-1}\right]_{ij}\left(M_{x}^{2}\right)E_{\phantom{i}x}^{i}E_{\phantom{j}x}^{j}\xrightarrow[]{p^2\rightarrow M_x^2} -iZ_x^{-1}\left(p^2-M^2_x\right)\;.
\end{equation}
Thus, noting that the projection vectors do not depend on $p^2$ we obtain $Z_x$ from
\begin{equation}\label{eq:Zq_inv}
  Z_x^{-1}=  E_{\phantom{i}x}^{i}E_{\phantom{j}x}^{j}\left\{i\partial_{p^2}\left[G^{-1}\right]_{ij}\right\}_{p^2=M^2_x} = 1-E_{\phantom{i}x}^{i}E_{\phantom{j}x}^{j}\left\{\partial_{p^2}\Delta\Sigma_{ij}\right\}_{p^2=M^2_x}\;.
\end{equation}

\section{General loop expansion and perturbative strategy}\label{sec:conventions}

In this section we aim to organise the perturbative strategy to obtain the one loop corrections to observables. To do so we construct the loop expansions by consistently truncating in powers of $\hbar$. It will be convenient to use as expansion parameter the quantity $\varepsilon\equiv \hbar/(4\pi)^2$. 

We start with some general considerations, assuming that a renormalisation scheme has been fixed and that, in principle, we have calculated a loop expansion for any observable as a function of a given set of input parameters for the theory. The number of such input parameters is equal to the number of running couplings appearing in the tree level Lagrangian. However, one may want to use a different set of input parameters, for example physical observables that are measured experimentally, to replace such Lagrangian parameters. With that choice then the Lagrangian parameters will be functions of such (observable) inputs. A more general and convenient approach may be to choose a mixture of observables and Lagrangian parameters, or even other theoretical parameters such as VEVs, as input.

 To avoid choosing, a priori, a particular set of inputs, we adopt a perturbative strategy to obtain radiatively corrected relations among the various parameters. We formally loop expand all parameters to allow for a free choice of inputs. To specify which parameters are input we can then simply set their correction terms to zero. An advantage of this procedure is that we obtain a linear system of constraints that we can analyse to decide what are the available choices of sets of input parameters.\footnote{Note that, in principle, not all choices allow to invert back the system to obtain the Lagrangian parameters so this procedure automatically displays the choices that are valid.} This flexibility is particularly well suited for the automation of the calculation of higher order corrections in general purpose tools to scan the parameter space of general scalar extensions of the SM~\cite{ScannerS,Coimbra:2013qq}.

To organise the system, let us denote collectively the set
of Lagrangian parameters, VEVs and observables that we want to expand by
$Q_{A}$. For example $A$ could run over the elements of the list $\left\{ C_{\lambda},v_{i},m_{x}^{2},\ldots\right\} $ with $C_\lambda$ representing the set of all Lagrangian parameters with\footnote{For example, in the $\Lambda$-basis, it is the restriction of the list $\{\Lambda_T^{\ldots},M^{IJ},Y^{IJk},G^{aij},G^{abc}\}$ to the set of independent parameters after the symmetries of the tensors are taken into account.} $\lambda=1,\ldots,N_\lambda$; $v_i$ the VEVs and $m_x^2$ the physical masses.
We assume that all quantities (for concreteness, renormalised in the $\overline{\rm MS}$-scheme) are loop expanded as 
\begin{equation}
Q_{A}=\sum_{n=0}^{+\infty}\varepsilon^{n}Q_{A}^{(n)}\; ,
\end{equation}
 where $Q_{A}^{(n)}$ is the $n$-loop order correction ($Q_{A}^{(0)}$ is the tree level quantity). Furthermore we assume that there is a set of constraints. Those can be, for example, the definition of observables (such as pole masses) or theoretical conditions such as the vacuum (or minimum) conditions. In general,
the system of constraints  is  represented by 
\begin{equation}
C_{\Gamma}\left(Q\right)=0 \; ,
\end{equation}
where $C_{\Gamma}$ denotes a constraint, with the index $\Gamma$ running over all the available constraints, and we suppress the index of $Q_{A}$ in arguments of functions for a lighter notation. Typically the constraints are also loop expanded as: 
\begin{equation}
\sum_{n=0}^{+\infty}\varepsilon^{n}C_{\Gamma}^{(n)}\left(Q\right)=0 \; .
\end{equation}
Expanding up to linear order we obtain 
\begin{eqnarray}
\Leftrightarrow C_{\Gamma}^{(0)}\left(Q^{(0)}\right)+\varepsilon\left[\partial_{B}C_{\Gamma}^{(0)}\left(Q^{(0)}\right)Q^{(1)B}+C_{\Gamma}^{(1)}\left(Q^{(0)}\right)\right]+\ldots & = & 0 \; .
\end{eqnarray}
 Equating order by order we obtain (for the first two orders)
\begin{equation}\label{eq:GenExpansionConstraints}
\begin{cases}
C_{\Gamma}^{(0)}\left(Q^{(0)}\right)=0 & \qquad n=0\\
\partial_{B}C_{\Gamma}^{(0)}\left(Q^{(0)}\right)Q^{(1)B}=-C_{\Gamma}^{(1)}\left(Q^{(0)}\right) & \qquad n=1 \;,
\end{cases} \; 
\end{equation}
so once we solve the tree level constraints for $Q^{(0)}_A$ we obtain a non-homogeneous linear system of constraints for the one-loop corrections $Q^{(1)}_A$. 

For concreteness we now turn to the problem we want to solve. First, we wish to compute the one-loop corrections to the constraints that define the vacuum state and the physical masses in the scalar sector. With this we will obtain relations among the parameters of the scalar sector of the theory. We assume that the gauge and fermion sector couplings are input parameters that are known at the renormalisation scale $\mu$. As already stated we formally allow all scalar couplings, all VEVs and all parameters defining the mass eigenstates to be loop expanded. Later we will decide, on a model by model basis, which parameters of the scalar sector are input. The set of constraints to impose are 
\begin{equation}
\begin{cases}
\partial_{i}V_{{\rm eff}}\left(v_{k},C_{\lambda}\right)=0\\
\left[M_x^{2}\delta_{ij}-\partial_{ij}^{2}V_{{\rm eff}}\left(v_{k},C_{\lambda}\right)-\Delta\Sigma_{ij}\left(M_x^{2},C_{\lambda}\right)\right]E_{\phantom{i}x}^{j}=0\;,
\end{cases}
\end{equation}
which are, respectively, the minimum conditions (tadpole equations) and the pole equations, which define the physical mass eigenstates -- see also Eq.~\eqref{eq:GenPoleConditions}.
 The parameters of the theory that can, in principle, be expanded are
\begin{equation}\label{eq:GenExpansionParameters}
\begin{cases}
C_{\lambda}=C_{\lambda}^{(0)}+\varepsilon C_{\lambda}^{(1)}+\ldots\\
v_{k}=v_{k}^{(0)}+\varepsilon v_{k}^{(1)}+\ldots\\
E_{\phantom{i}x}^{j}=E_{\phantom{(0)i}x}^{(0)j}+\varepsilon E_{\phantom{(0)i}x}^{(1)j}+\ldots\\
m_x^{2}=m_x^{(0)2}+\varepsilon m_x^{(1)2}+\ldots\\
\Gamma_x=0+\varepsilon\Gamma_x^{(1)}+\ldots
\end{cases}
\end{equation}
 where we have noted that the width is always zero at zeroth order so the Leading Order (LO) results appears only at first order. We also note that 
\begin{equation}\label{eq:GenExpansionVeff_Sigma}
\begin{cases}
V_{{\rm eff}}\left(v_{k},C_{\lambda}\right)=V^{(0)}\left(v_{k},C_{\lambda}\right)+\varepsilon V^{(1)}\left(v_{k},C_{\lambda}\right)+\ldots\\
\Delta\Sigma_{ij}\left(M_x^{2},C_{\lambda}\right)=0+\varepsilon\Delta\Sigma_{ij}^{(1)}\left(M_x^{2},C_{\lambda}\right)+\ldots
\end{cases}
\end{equation}
where $V^{(n)}$ is the $n$-loop effective potential and the self energy series only starts at first order. Using the general expansion, Eq.~\eqref{eq:GenExpansionConstraints} and inserting the expansions, Eqs.~\eqref{eq:GenExpansionParameters} and~\eqref{eq:GenExpansionVeff_Sigma}, and assuming, without loss of generality, that we are in a field basis such that the tree level scalar eigen-states are aligned along each field direction, i.e. $E_{\phantom{(0)i}x}^{(0)j}=\delta_x^j$, we obtain the tree level conditions
\begin{equation}
\begin{cases}
\left[\partial_{i}V^{(0)}\right]_{{\rm tree}}=0\\
m_x^{(0)2}\delta_{ix}-\left[\partial_{ix}^{2}V^{(0)}\right]_{{\rm tree}}=0
\end{cases}
\end{equation}
 and a linear system for the one-loop corrections
\begin{equation}
\begin{cases}
\left[\partial_{ij}^{2}V^{(0)}\right]_{{\rm tree}}v^{(1)j}+\left[\partial_{i\lambda}^{2}V^{(0)}\right]_{{\rm tree}}C^{(1)\lambda}=-\left[\partial_{i}V^{(1)}\right]_{{\rm tree}}\\
\left(m_x^{(0)2}\delta_{ij}-\left[\partial_{ij}^{2}V^{(0)}\right]_{{\rm tree}}\right)E_{\phantom{(0)i}x}^{(1)j}+\left(m_x^{(1)2}-i\Gamma_x^{(1)}m_x^{(0)2}\right)E^{(0)}_{ix}\\
-\left(\left[\partial_{ijk}^{3}V^{(0)}\right]_{{\rm tree}}v^{(1)k}+\left[\partial_{ij\lambda}^{3}V^{(0)}\right]_{{\rm tree}}C^{(1)\lambda}\right)E^{(0)j}_{\phantom{(0)j}x}=\left[\partial_{ij}^{2}V^{(1)}+\Delta\Sigma_{ij}^{(1)}\right]_{{\rm tree}}E^{(0)j}_{\phantom{(0)j}x} \; .
\end{cases}
\end{equation}
Here we use the notation $[\ldots]_{\rm tree}$ to denote a quantity evaluated with tree level arguments and $\Delta\Sigma_{ij}^{(1)}$ is evaluated at $s=m_x^{(0)2}$. Taking the real and imaginary parts of the second constraint, and noting that, in the tree level basis, the tree level mass squared matrix is diagonal, we obtain the final result
\begin{equation}\label{eq:PoleEqs1loop}
\begin{cases}
m_{i}^{2(0)}v_{i}^{(1)}+\left[\partial_{i\lambda}^{2}V^{(0)}\right]_{{\rm tree}}C^{(1)\lambda}=-\left[\partial_{i}V^{(1)}\right]_{{\rm tree}}\\
\left(m_x^{2(0)}-m_{i}^{2(0)}\right)E_{\phantom{(0)}ix}^{(1)}+m_x^{2(1)}\delta_{ix}-\left[\partial_{ixk}^{3}V^{(0)}\right]_{{\rm tree}}v^{(1)k}-\left[\partial_{ix\lambda}^{3}V^{(0)}\right]_{{\rm tree}}C^{(1)\lambda}\\
=\left[\partial_{ix}^{2}V^{(1)}+\Re\left(\Delta\Sigma_{ix}^{(1)}\right)\right]_{{\rm tree}}\\
\Gamma_x^{(1)}m_x^{(0)2}\delta_{ix}=-\Im\left[\Delta\Sigma_{ix}^{(1)}\right]_{{\rm tree}}\;.
\end{cases}
\end{equation}
To this, we can add the conditions that the eigenstates are normalised which, up to one-loop order, translates to 
\begin{equation}\label{Eq:choice mixing_offdiag}
E^i_xE_{ix}=1\Rightarrow 1+2\varepsilon E^{(1)}_{xx}+O(\varepsilon^2)=1\Rightarrow E^{(1)}_{xx}=0\; .
\end{equation}

Now that we wrote the general system, Eq.~\eqref{eq:PoleEqs1loop}, we discuss some possible choices of inputs. If, for example, one chooses to take as input parameters the set of running Lagrangian parameters $C_\lambda$,  we set  $C_{\lambda}^{(i>0)}\equiv 0$. Then the one-loop shifts of all other quantities are computed using the system in Eq.~\eqref{eq:PoleEqs1loop}. Within that choice, using the one-loop tadpole equation in the first line of Eq.~\eqref{eq:PoleEqs1loop} we obtain directly that, for consistency, for states that are massless at tree level, the first derivative of the one-loop effective potential evaluated at the tree level couplings must be zero and the corresponding VEV shift remains undetermined. Denoting the field space directions associated with the massless states by the sub-indices $\epsilon_1,\epsilon_2,\ldots$ and the ones associated with massive states with barred indices $\bar{i},\bar{j}\ldots$, then the remaining VEVs are obtained from
\begin{equation}
v_{\bar{i}}^{(1)}=-\left(m_{\bar{i}}^{2(0)}\right)^{-1}\left[\partial_{\bar{i}}V^{(1)}\right]_{{\rm tree}}\; .
\end{equation}
All that is left in this case is to solve the pole equations, Eq.~\eqref{eq:PoleEqs1loop} with $C_{\lambda}^{(1)}\equiv 0$. 

Other possible choices consist of perturbative inversions of the one-loop relations in Eq.~\eqref{eq:PoleEqs1loop}.  In our examples in Sect.~\ref{sec:applications} we will choose some input parameters to be physical quantities (such as masses) and others such as the Higgs VEV, mixing matrix elements and a few Lagrangian parameters. This is convenient, for example, to fix the one-loop Higgs mass to the experimental value of $125$~GeV and the Higgs VEV to $246$~GeV. Then the one-loop shifts of the remaining parameters are computed, ensuring that the relations among all parameters are correct to one-loop order.

A particularly choice that is useful, is one that decouples directly the corrections to the mass eigenstates states. Taking the anti-symmetric part of the second condition in Eq.~\eqref{eq:PoleEqs1loop} for $i\neq x$ we obtain
\begin{equation}\label{eq:one_loop_mix_anti_sym}
E_{\phantom{(0)}ix}^{(1)}
=-E_{\phantom{(0)}xi}^{(1)}+\frac{\Re\left[\Delta\Sigma_{ix}^{(1)}-\Delta\Sigma_{xi}^{(1)}\right]_{{\rm tree}}}{m_x^{2(0)}-m_{i}^{2(0)}} \; .
\end{equation}
So, assuming that the system allows the choice $E_{\phantom{(0)}ix}^{(1)}=0$ with $i>x$, we get the solution for all the corrections to the mass eigen-state expansions.

Finally, another set of quantities that we will need are the wave function renormalisation factors. Expanding Eq.~\eqref{eq:Zq_inv} perturbatively  we find
\begin{equation}\label{eq:Z_pert_expand}
Z_x^{-1}=  1-\varepsilon \left[\partial_{p^2}\Delta\Sigma_{xx}\right]_{\rm tree}+O\left(\varepsilon^2\right)\;.
\end{equation}

\subsection{Coleman-Weinberg potential and self energies}\label{subsec:Coleman-Weinberg}

The quantities we will need to evaluate in Eq.~\eqref{eq:PoleEqs1loop} are: the Coleman-Weinberg potential, its first and second derivatives and the variation in the self-energy functions. In the pole conditions one could, equivalently, simply compute the full self-energy functions. However, in Eq.~\eqref{eq:GenPoleConditions} we have written the result in terms of the effective potential to separate out the $p^2$-independent part and the $p^2$-dependence. This is useful to connect to the $p^2\rightarrow 0$ approximation, which can be used if the dominant contributions to the radiative corrections are from heavy particles in the loops~\cite{Camargo-Molina:2016moz}. In that limit the effective potential encodes all the necessary information. 

The general one-loop effective potential in the $\overline{\rm MS}$ scheme is given by the Coleman-Weinberg potential. Recently~\cite{Camargo-Molina:2016moz} we have analysed the Coleman-Weinberg potential and obtained a closed form master formula for its one-loop derivatives with any number of external scalar field legs for a general theory as described in Sect.~\ref{sec:DefsNotation}. Here we only review the two expressions that we need, i.e. the first and the second derivatives of the effective potential, respectively,
\begin{equation}\label{eq:GenFirstDerivs}
\partial_iV^{(1)}=\sum_{T}\tfrac{(-1)^{2s_{T}}(1+2s_{T})}{2}m_{(T)a}^{2}\lambda_{(T)a{\phantom a}i}^{\phantom{(T)a}a}\left(\logbar m_{(T)a}^{2}-k_{T}+\frac{1}{2}\right)
\end{equation}
and
\begin{multline}\label{eq:GenSecondDerivs}
\partial_{ij}^2V^{(1)}=\sum_{T}\tfrac{(-1)^{2s_{T}}(1+2s_{T})}{2}{\rm S}_{\{ij\}}\left[\lambda_{(T)i}^{ab}\lambda_{(T)j}^{ba}\left(f_{(T)ab}^{(1)}-k_{T}+\frac{1}{2}\right)+\right.\\\left.+\lambda_{(T)aij}^{a}m_{(T)a}^{2}\left(\logbar m_{(T)a}^{2}-k_{T}+\frac{1}{2}\right)\right]\; .
\end{multline}
Here $\logbar(x)\equiv \log(x/\mu^2)$, with $\mu$ being the renormalisation scale; $s_T=0,\frac{1}{2}$ or $1$ is the spin of the field of type $T$ and the $\lambda_{(T)}^{\ldots}$ couplings have been defined in Eq.~\eqref{Eq:lambda_basis}; ${\rm S}_{\{ij\}}$ denotes symmetrisation of the indices; and 
\begin{equation}\label{eq:fnT-tensors}
f^{(1)}_{(T)a_{1}\ldots a_{N}}\equiv \sum_{x=1}^{N}\dfrac{m^2_{(T)a_x}\logbar {m^2_{(T)}}_{a_x}}{\prod_{y\neq x}\left(m^2_{(T)a_x}-m^2_{(T)a_y}\right)}\; .
\end{equation}   
Observe that the latin indices $a,b,\ldots$ in the $\lambda_{(T)}^{\ldots}$ tensors are to be replaced by scalar, fermionic or vector indices respectively according to $T$. The constant $k_T$ depends on the renormalisation scheme (for $\overline{\rm MS}$ it is $3/2$ for scalars and fermions and $5/6$ for vector bosons).  

Regarding the self-energies, they have been computed in~\cite{Martin:2003it}. Here we present the variation that we need, which is ($s\equiv p^2$) 
\begin{eqnarray}\label{eq:DeltaSigma}
\Delta\Sigma^{(1)}_{ij}(s) & = & \dfrac{1}{2}\lambda^{kl}_{(S)i}\lambda^{kl}_{(S)j}\Delta B_{SS}\left(m_{k}^{2},m_{l}^{2}\right)+\Re\left[y^{KL}_{\phantom{KL}i}y_{KLj}^\star\right]\Delta B_{FF}(m_{K}^{2},m_{L}^{2})+\nonumber \\
&&+\Re\left[y^{KL}_{\phantom{KL}i}m_{KK'}^{\star}m_{LL'}^{\star}y^{K'L'}_{\phantom{K'L'}j}\right]\Delta B_{\bar{F}\bar{F}}(m_{K}^{2},m_{L}^{2}) \\
&&+g^{ak}_{\phantom{ak}i}g^{ak}_{\phantom{ak}j}\Delta B_{SV}\left(m_{k}^{2},m_{a}^{2}\right)+\frac{1}{2}\lambda^{ab}_{(G)i}\lambda^{ab}_{(G)j}\Delta B_{VV}\left(m_{a}^{2},m_{b}^{2}\right) \; .\nonumber
\end{eqnarray}
The various loop function variations can be obtained directly from the results in~\cite{Martin:2003it,Martin:2003qz} and are provided in appendix~\ref{app:LoopFuncs}.

\subsection{Infrared behaviour}\label{subsec:IRbehaviour}

It is well known~\cite{Elias-Miro:2014pca,Martin:2014bca} that the derivatives of second and higher orders of the (one-loop) Coleman-Weinberg potential can contain infrared divergences originating from the massless states running in the loops. However this is not a problem if all the $p^2$ dependent contributions are included because all such infrared divergences must cancel out. 

In this section we verify this general cancellation explicitly and write the final result in a manifestly regular form suitable for numerical evaluation. First let us introduce an infrared regulator mass squared scale, $\epsilon$. The second derivatives of the effective potential can be split as 
\begin{equation}
\partial^2_{ij}V^{(1)}=\partial^2_{ij}V^{(1)}_{\rm finite}+\partial^2_{ij}V^{(1)}_{\rm IR}
\end{equation}
where the second term contains the contributions with internal sums over the indices $a,b$ corresponding to two internal massless states -- see Eq.~\eqref{eq:GenSecondDerivs}. We recall that, in our notation, indices corresponding to massless state components are denoted by $\epsilon_1,\epsilon_2,\ldots$ when the type of the field, $T$, is not specified. Whenever $T$ is specified, we use the indices $\epsilon_1,\epsilon_2,\ldots$ for scalar indices, $E_1,E_2,\ldots$ for fermionic indices and $e_1,e_2,\ldots$ for vector indices. With this notation, the IR-divergent piece is
\begin{eqnarray}
\partial_{ij}^{2}V_{\rm IR}^{(1)} & = & \sum_{T}\tfrac{(-1)^{2s_{T}}(1+2s_{T})}{2}{\rm S}_{\{ij\}}\hspace{-1mm}\left[\lambda_{(T)i}^{\epsilon_{1}\epsilon_{2}}\lambda_{(T)j}^{\epsilon_{2}\epsilon_{1}}\hspace{-1mm}\left(f_{(T)\epsilon_{1}\epsilon_{2}}^{(1)}-k_{T}+\tfrac{1}{2}\right)+\lambda_{(T)\epsilon_1 ij}^{\epsilon_1}\epsilon\left(\logbar\epsilon-k_{T}+\tfrac{1}{2}\right)\right]\nonumber\\
 & = & \logbar\epsilon\left\{ \frac{1}{2}\lambda_{(S)\phantom{,}i}^{\epsilon_{1}\epsilon_{2}}\lambda_{(S)\epsilon_{2}\epsilon_{1}j}-\Re\left[\lambda_{(F)\phantom{,}i}^{E_{1}E_{2}}\lambda_{(F)E_{2}E_{1}j}\right]+\frac{3}{2}\lambda_{(G)\phantom{,}i}^{e_{1}e_{2}}\lambda_{(G)e_{2}e_{1}j}\right\} +\nonumber\\
&&\hspace{47mm}+\sum_{T}\tfrac{(-1)^{2s_{T}}(1+2s_{T})}{4}\left[\lambda_{(T)i}^{\epsilon_{1}\epsilon_{2}}\lambda_{(T)\epsilon_{2}\epsilon_{1}j}+c.c.\right]\left(\frac{3}{2}-k_{T}\right)\nonumber\\
 & \equiv & \partial_{ij}^{2}V_{\rm IR, div}^{(1)}+\partial_{ij}^{2}V_{\rm IR, finite}^{(1)}\;,
\end{eqnarray}
where, on the second line, we have series expanded in the cutoff, $\epsilon$, and kept only the divergent and constant terms, respectively denoted by $\partial_{ij}^{2}V_{\rm IR, div}^{(1)}$ and $\partial_{ij}^{2}V_{\rm IR, finite}^{(1)}$. In the divergent term, $\partial_{ij}^{2}V_{\rm IR, div}^{(1)}$, we have explicitly expanded over spins. 

Moving on to the self-energies, we define a similar split 
\begin{equation}
\Delta\Sigma^{(1)}_{ij}(s)=\Delta{\Sigma}^{(1)}_{ij,{\rm finite}}(s)+\Delta\Sigma_{ij,{\rm IR}}^{(1)}(s)
\end{equation}
where now 
\begin{eqnarray}
\Delta\Sigma_{ij,{\rm IR}}^{(1)}(s) & = & \dfrac{1}{2}\lambda^{\epsilon_{1}\epsilon_{2}}_{(S)\phantom{,}i}\lambda_{(S)\epsilon_{1}\epsilon_{2}j}\Delta B_{SS}\left(\epsilon,\epsilon\right)+\Re\left[y^{E_{1}E_{2}}_{\phantom{E_{1}E_{2}}i}y_{E_{1}E_{2}j}^{\star\phantom{KL}}\right]\Delta B_{FF}(\epsilon,\epsilon)+\nonumber\\
&&+\Re\left[y^{E_{1}E_{2}}_{\phantom{E_{1}E_{2}}i}m_{E_{1}K'}^{\star}m_{E_{2}L'}^{\star}y^{K'L'}_{\phantom{K'L'}j}\right]\Delta B_{\bar{F}\bar{F}}(\epsilon,\epsilon)+\\
 &  & +g^{e_{1}\epsilon_{2}}_{\phantom{e_{1}e_{2}}i}g_{e_{1}\epsilon_{2}j}\Delta B_{SV}\left(\epsilon,\epsilon\right)+\frac{1}{2}\lambda^{e_{1}e_{2}}_{(G)\phantom{,}i}\lambda_{(G)e_{1}e_{2}j}\Delta B_{VV}\left(\epsilon,\epsilon\right)\;. \nonumber
\end{eqnarray}
Finally, using the fact that, in the mass-squared eigenbasis, $m_{IJ}$ only has non-zero elements between states $I,J$ with the same mass, using Eq.~\eqref{Eq:gammas_to_ys} in appendix~\ref{app:identities} and Eqs.~\eqref{Eq:Bssepsilonepsilon}, \eqref{Eq:BFFepsilonepsilon} and~\eqref{Eq:BVVepsilonepsilon} in appendix~\ref{app:LoopFuncs} we find that 
\begin{eqnarray}
\Delta\Sigma_{ij,{\rm IR}}^{(1)}(s) & = & -\partial^2_{ij}V_{\rm IR,div}^{(1)}+\left\{\dfrac{1}{2}\lambda^{\epsilon_{1}\epsilon_{2}}_{(S)\phantom{,}i}\lambda_{(S)\epsilon_{1}\epsilon_{2}j}-\Re\left[\lambda_{(F)\phantom{,}i}^{E_{1}E_{2}}\lambda_{(F)E_{2}E_{1}j}\right]\right\}\left(\logbar s-2-i\pi\right)\nonumber\\
 &  & +\frac{3}{2}\lambda^{e_{1}e_{2}}_{(G)\phantom{,}i}\lambda_{(G)e_{1}e_{2}j}\left(\logbar s-\tfrac{3}{2}-i\pi\right)+\nonumber\\
&&+g^{e_{1}\epsilon_{2}}_{\phantom{e_{1}e_{2}}i}g_{e_{1}\epsilon_{2}j}\Delta B_{SV}\left(0,0\right)+\Re\left[y^{E_{1}E_{2}}_{\phantom{E_{1}E_{2}}i}y_{E_{1}E_{2}j}^\star\right]\Delta B_{FF}(0,0)\quad\nonumber\\
 & \equiv & -\partial^2_{ij}V_{\rm IR,div}^{(1)}+\Delta\Sigma_{ij,{\rm IR,finite}}^{(1)}(s)\; .
\end{eqnarray}
Therefore the divergent pieces cancel precisely. The final, explicitly finite, result for the full one loop contributions appearing in the pole equations is then 
\begin{equation}\label{eq:IR_safe_expression}
\partial_{ij}^{2}V^{(1)}+\Delta\Sigma_{ij}^{(1)}(s)=\partial_{ij}^{2}V^{(1)}_{\rm finite}+\partial_{ij}^{2}V_{\rm IR, finite}^{(1)}+\Delta\Sigma_{ij,{\rm finite}}^{(1)}(s)+\Delta\Sigma_{ij,{\rm IR, finite}}^{(1)}(s)\; .
\end{equation}

\section{Application to general scalar singlet extensions of the SM}
\label{sec:applications}
In this section we will apply the results of the previous section to the most General scalar singlet extension of the SM (GxSM) and then we specialise to a real (RxSM) and a complex (CxSM) singlet extensions. For simplicity we work in the $\overline{\rm MS}$ scheme, so we replace the $k_T$ by their numerical values.
 
\subsection{Definition of the GxSM}
The most general scalar singlet extension of the SM is obtained by adding to the Lagrangian a set $S_k$ ($k=1,\ldots,N_S$) of real scalar hypercharge zero singlet fields with a general renormalisable scalar potential. The Lagrangian density of the model for the interaction terms is then
\begin{equation}
-\mathcal{L}_{\rm int}=-\mathcal{L}_{\rm int,SM}+ \Delta (S)H^{\dagger}H+V(S)
\end{equation}
where $\Delta(S)$ and $V(S)$ are polynomials that are, respectively, up to quadratic  and quartic in the fields $S_k$ (without constant terms) and $H$ is the SM Higgs doublet. In this framework the full scalar potential is then 
\begin{equation}
V_{\rm GxSM}=\dfrac{m^2}{2}H^\dagger H+\dfrac{\lambda}{4}(H^\dagger H)^2+ H^{\dagger}H\Delta (S)+V(S)\; .
\end{equation}
One or more of the singlet fields can mix with the Higgs boson provided that $\partial_k\Delta(v_i)\neq 0$ for at least one value of $k$ at the electroweak symmetry breaking vacuum with a choice of VEVs $v,v_k$ such that  
\begin{equation}\label{eq:vacua_GxSM}
H=\dfrac{1}{\sqrt{2}}\left(\begin{array}{c} G^+ \\
    v+h+iG^0\end{array}\right) \quad \mbox{and} \quad
S_k=v_k+s_k \;.
\end{equation}
Here $h$ is the SM Higgs field fluctuation, $G_0,G^+$ are the Goldstones and $s_k$ are the singlet field fluctuations around the vacuum.
The new scalar singlet fields, $S_k$, do not couple directly to other SM fields. As a consequence, the tree level coupling of the scalar mass eigenstates that are a mixture of singlet field fluctuations, $s_k$, with the Higgs boson fluctuation, $h$, to the other SM particles is simply scaled by a mixing factor (compared with the Higgs couplings in the SM). We note that, at tree level, the Higgs field fluctuation is decomposed in terms of scalar mass eigenstates as (see Eq.~\eqref{Eq:Rot-fields})
\begin{equation}
h=\left[O_{(S)}\right]_{1j}R^j\equiv \kappa_{j}R^j
\end{equation}
where we have ordered the set of scalar fields after the VEVs shift as \[\phi^T=(h,s_1,s_2,\ldots,G^0,\Re[G^+],\Im[G^+])\; .\] Here $\kappa_j$ is the scaling factor to apply to the SM coupling of an SM-like Higgs of the same mass as the state $R_j$, to obtain the coupling of that state in the GxSM. Due to the orthogonality of the mixing matrix we have that, at tree level,
\begin{equation}\label{eq:tree_sum}
  \sum_j\kappa_j^2=1
\end{equation}
which means that the SM-like coupling is shared among the Higgs like states~\cite{Barger:2009me}. As a consequence the one-loop radiative corrections to the scalar mass eigenstates will contain some SM-like contributions suitably suppressed by the dilution factors $\kappa_j$ and also contributions exclusively due to the new scalar sector.

\subsection{NLO parameter shifts}
In this section we compute the NLO shifts of the parameters in the GxSM.
\subsubsection*{One-loop tadpoles}

The contributions to the tadpole conditions, Eqs.~\eqref{eq:GenFirstDerivs} all contain a coupling factor $\lambda_{(T)aai}$. For fermions and vector bosons we know that, for the GxSM, the couplings to the massive scalars are simply scaled by a $\kappa_j$ factor and the couplings to Goldstone bosons are precisely the same as in the SM. Thus, noting that from now on we no longer deal with space-time indices, if we use Greek indices $\alpha = 1,\ldots,4$  to denote the four scalar degrees of freedom in the SM and define a dilution tensor $D_i ^\alpha$ ($n_s$ is the number of non-Goldstone real scalars)
\begin{equation}
  D_i^\alpha =\begin{cases}
  \kappa_i & \alpha = 1 \;(h)\\
  \delta_{i}^{\alpha+n_s-1} & \alpha =2,3,4\; {\rm (goldstones)}
  \end{cases}
\end{equation}
then
\begin{eqnarray}\label{eq:SM_like_couplings}
y_{IJi}&=&D_i^\alpha y_{IJ\alpha}^{\rm SM}\nonumber\\
\lambda_{(T\neq S)abi}&=& D_i^\alpha\,\lambda_{(T)ab\alpha}^{\rm SM}\\
\lambda_{(T\neq S)abij}&=& D_i^\alpha D_j^\beta\,\lambda_{(T)ab\alpha \beta}^{\rm SM}\nonumber \\
g_{aij}&=& D_i^\alpha D_j^\beta\,g_{(T)a\alpha \beta}^{\rm SM} \nonumber\;,
\end{eqnarray}
where we have denoted the SM couplings on the right hand side with the superscript SM. Then one can check that (see also appendix~\ref{app:SMcontribs}) 
\begin{eqnarray}
\partial_iV^{(1)}&=&\tfrac{1}{2}\lambda_{(S)ki}^{k}m_{k}^{2}\left(\logbar m_{k}^{2}-1\right)+D_i^\alpha\sum_{T\neq S}\tfrac{(-1)^{2s_{T}}(1+2s_{T})}{2}m_{(T)a}^{2}\lambda_{(T)a{\phantom a}\alpha }^{\phantom{(T)a}a}\left[\logbar m_{(T)a}^{2}-k_{T}+\tfrac{1}{2}\right]\nonumber\\
&\simeq & \tfrac{1}{2}\lambda_{(S)ki}^{k}m_{k}^{2}\left(\logbar m_{k}^{2}-1\right)-6\kappa_im_{t}^{2}y_t^2v\left(\logbar m_{t}^{2}-1\right)+\\
&&+2\kappa_i\frac{m_{W}^4}{v}\left(3\logbar m_{W}^{2}-1\right)+\kappa_i\frac{m_{Z}^4}{v}\left(3\logbar m_{Z}^{2}-1\right)\nonumber
\end{eqnarray}
where, in the last line, we have evaluated the result keeping only the dominant top quark contribution in the fermion sector and the electroweak vector boson contributions -- see appendix~\ref{app:SMcontribs}.

\subsubsection*{One-loop poles and wave function renormalisations}
From the pole equations, Eq.~\eqref{eq:PoleEqs1loop} and using Eq.~\eqref{eq:SM_like_couplings} one can show that 
\begin{eqnarray}\label{eq:ZfactorsGxSM}
  &&\partial_{ij}^{2}V^{(1)}+\Delta\Sigma_{ij}^{(1)} \\
  &=& \left[\partial_{ij}^{2}V^{(1)}+\Delta\Sigma_{ij}^{(1)}\right]_{\rm scalars}+D_i^\alpha D_j^\beta\sum_l \kappa_l^2\left[\partial_{\alpha\beta}^{2}V^{(1)}_{\rm SM}+\Delta\Sigma_{\alpha\beta,{\rm SM}}^{(1)}\right]_{(T\neq S),m_h^2=m_l^2}\nonumber\\
  & \simeq & \left[\partial_{ij}^{2}V^{(1)}+\Delta\Sigma_{ij}^{(1)}\right]_{\rm scalars}+D_i^\alpha D_j^\alpha \left[S^{(t)}_{\alpha}(s)+ S^{(g,1)}_{\alpha}(s)+\sum_l \kappa_l^2S^{(g,2)}_{\alpha}(m_l^2,s)\right]\; ,\nonumber
\end{eqnarray}
where we use the approximation with only the top quark contribution in the fermion sector and where the SM quantities are evaluated with the Higgs mass replaced with the mass $m_l^2$. The vectors $S_i^{(\ldots)}$ are defined in appendix ~\ref{app:SMcontribs}. The function $B_s(x,y)$ can be found in appendix~\ref{app:LoopFuncs}, Eq.~\eqref{eq:Bxy}. The scalar contributions can also be simplified using the IR safe expression, Eq.~\eqref{eq:IR_safe_expression},
\begin{multline}
  \label{eq:GxSMpolesScalars}
\hspace{-4mm}\left[\partial_{ij}^{2}V^{(1)}+\Delta\Sigma_{ij}^{(1)}\right]_{{\rm scalars}}=\frac{1}{2}\left[-\lambda_{(S)i}^{\bar{k}\bar{l}}\lambda_{(S)j}^{\bar{k}\bar{l}}B_s\left(m_{\bar{k}}^{2},m_{\bar{l}}^{2}\right)-2\lambda_{(S)i}^{\epsilon\bar{k}}\lambda_{(S)\epsilon j}^{\bar{k}}B_s\left(0,m_{\bar{k}}^{2}\right)+\right.\\
\left.\phantom{\frac{1}{2}}+\lambda_{(S)\bar{k}ij}^{\bar{k}}m_{\bar{k}}^{2}\left(\logbar m_{\bar{k}}^{2}-1\right)+\lambda_{(S)i}^{\epsilon_{1}\epsilon_{2}}\lambda_{(S)\epsilon_{1}\epsilon_{2}j}\left(\logbar s-2-i\pi\right)\right]\; ,
\end{multline}
where, again, the barred indices run only over eigenstates with a non-zero mass.
In practice we will be interested in the components $i,x$ such that $s=m_x^2$ so we have 
\begin{eqnarray}
  &&\partial_{ix}^{2}V^{(1)}+\Delta\Sigma_{ix}^{(1)}\\
  & \simeq &  \left[\partial_{ix}^{2}V^{(1)}+\Delta\Sigma_{ix}^{(1)}\right]_{\rm scalars}+D_i^\alpha D_x^\alpha \left[S^{(t)}_{\alpha}(m_x^2)+ S^{(g,1)}_{\alpha}(m_x^2)+\sum_l \kappa_l^2S^{(g,2)}_{\alpha}(m_l^2,m_x^2)\right]\;. \nonumber
\end{eqnarray}

The other quantity that we will use are the one-loop wave function renormalisation factors for the massive states given by
\begin{eqnarray}\label{eq:Zqminus1}
  &&Z_x^{-1}-1\nonumber \\
  &=& -\varepsilon \left\{\partial_{s}\Delta\Sigma_{xx}\right\}_{s=M^2_x}+O\left(\varepsilon^2\right)\;\\
  &\simeq& -\varepsilon \left\{\partial_{s}\left[\Delta\Sigma_{xx}^{(1)}\right]_{\rm scalars}+\kappa_x^2 \left[\partial_sS^{(t)}_{h}(s)+ \partial_sS^{(g,1)}_{h}(s)+\sum_l \kappa_l^2\partial_sS^{(g,2)}_{h}(m_l^2,s)\right]\right\}_{s=m^2_x} \; , \nonumber
\end{eqnarray}
which, we can check, also involves $\partial_{s}B_s$. The term that will be most relevant is
\begin{equation}
\partial_{s}\left[\Delta\Sigma_{xx}^{(1)}\right]_{\rm scalars} = -\frac{1}{2}\lambda_{(S)x}^{\bar{k}\bar{l}}\lambda_{(S)x}^{\bar{k}\bar{l}}\partial_sB_s\left(m_{\bar{k}}^{2},m_{\bar{l}}^{2}\right)-\lambda_{(S)x}^{\epsilon\bar{k}}\lambda_{(S)\epsilon x}^{\bar{k}}\partial_sB_s\left(0,m_{\bar{k}}^{2}\right)+\frac{1}{2s}\lambda_{(S)x}^{\epsilon_{1}\epsilon_{2}}\lambda_{(S)\epsilon_{1}\epsilon_{2}x}
\end{equation}
evaluated at $s=m_x^2$.

With all these ingredients, we will specialise these formulas in Sect.~\ref{sec:Results} to obtain the parameter shifts in particular scalar singlet models.

\subsubsection*{Corrections to mixing sums}

In the GxSM, generically, there is a mixing of the SM Higgs field fluctuation with singlet fields. This typically results, at tree level, in a block $n$ by $n$ mixing matrix with $n$ the number of non-dark scalar mass eigenstates with the tree level sum rule, Eq.~\eqref{eq:tree_sum}, for the suppression factors of each scalar $x$ to other SM particles. If we denote the tree level suppression factor by  $\kappa_x^{(0)}$, at one loop using Eq.~\eqref{eq:GenExpansionParameters}, the one loop mass eigenstates in the gauge basis are (with $j,x$ running over the mass eigenstates)
\begin{equation}
  E^{i}_{\phantom{i}x}=\left[O_S\right]^i_{\phantom{i}j}\left(\delta^j_x+\varepsilon E^{(1)j}_{\phantom{(1)i}x}\right)\Rightarrow \kappa_x=\kappa^{(0)}_x+\varepsilon \kappa^{(0)}_jE^{(1)j}_{\phantom{(1)i}x}
\end{equation}
where now we denote by $\kappa_x$ the one loop corrected mixing factor. From this, and using Eq.~\eqref{eq:one_loop_mix_anti_sym}, we obtain the order $\varepsilon$ correction to this sum, which is given by
\begin{eqnarray}\label{eq:sum_rule_violation}
  \sum_{x}\kappa_x^2-1&=&\frac{\varepsilon}{2} \kappa^{(0)j}\kappa^{(0)x}\left(E^{(1)}_{\phantom{(1)}jx}+E^{(1)}_{\phantom{(1)}xj}\right)+O(\varepsilon^2)\\
  &=& \frac{\varepsilon}{2}\sum_{j\neq x} \kappa^{(0)j}\kappa^{(0)x}\frac{\Re\left[\Delta\Sigma_{jx}^{(1)}-\Delta\Sigma_{xj}^{(1)}\right]_{{\rm tree}}}{m_x^{2(0)}-m_{j}^{2(0)}}+O(\varepsilon^2)\; .\nonumber
\end{eqnarray}
This result is in fact independent of our choice of one-loop input parameters within the choice we made to normalise the mass eigenstates such that $E^{(1)}_{xx}=0$.
Thus, though we will evaluate Eq.~\eqref{eq:sum_rule_violation} in a specific scheme, the result is fixed within our class of schemes.

In our numerical results we will be interested in assessing the importance of the one loop corrections to the parameters of the theory. The quantity in Eq.~\eqref{eq:sum_rule_violation} is a good one to test the importance of these corrections as it is a shift of a tree level value that is $1$, and that does not depend on other choices within our class of schemes. Finally, this quantity is also of interest because it will contribute to the NLO sum rule that is expected to exist among the effective couplings of the mixing Higgs bosons to SM particles such as to preserve unitarity. A complete computation of such effective couplings is, however, beyond the scope of this study.

\subsection{NLO gluon fusion cross section}\label{sec:gluon_fusion_xs}
In Sect.~\ref{sec:Results}, we will evaluate the NLO electroweak corrections to the SM-like Higgs production cross-section in the gluon fusion channel that are due to the new scalar sector couplings. The current collider data already sets the suppression factor for the SM-like Higgs to be very close to unity and the suppression factors of the other new Higgs bosons to be very close to zero, i.e. $\kappa_h^2\sim 1$ and $\kappa_{i\neq h}^2\ll 1$. Therefore we will focus on this limit by systematically dropping terms that are suppressed by $\kappa_{i\neq h}$ or higher powers. Furthermore, using the standard assumption of factorisation of the QCD higher order corrections (see for example~\cite{Actis:2008ug}), we focus only on the NLO electroweak corrections. The NLO amplitude for gluon fusion in the GxSM is of the form:
\begin{equation}
A_{ggF}^{(NLO)}=\varepsilon\sqrt{Z_{h}}\left[\kappa_{h}A_{ggF}^{(LO)}+\varepsilon\left(\sum_{i}\kappa_{i}^{2}\kappa_{h}A_{iff}^{(NLO)}+\sum_{ij}\kappa_{i}\kappa_{j}\lambda_{ijh}A_{ijf}^{(NLO)}+\kappa_{h}A_{EW}^{(NLO)}\right)\right] 
\end{equation}
where $Z_h$ is the SM-like Higgs wave function renormalisation factor and the four amplitudes factors are (see also Fig.~\ref{fig:Feyn_diag_ggF}):
\begin{figure}[t]
\centering
\includegraphics[scale=0.4,trim=0 0 0 0,clip=true]{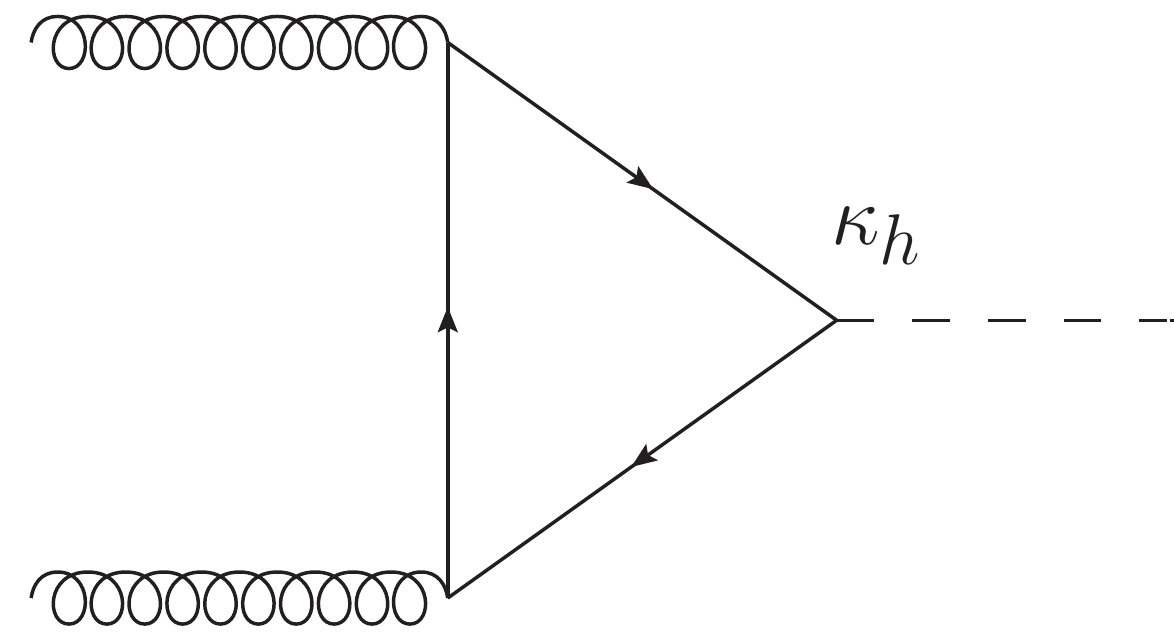}\hspace{4mm}\includegraphics[scale=0.4,trim=0 0 0 0,clip=true]{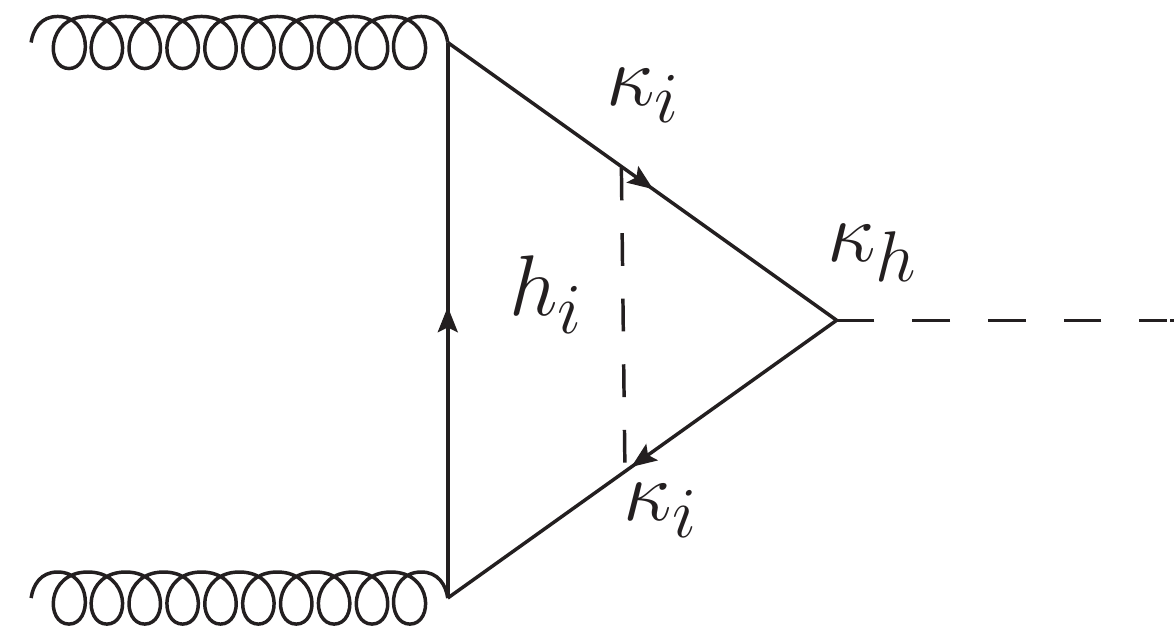}\hspace{4mm}\includegraphics[scale=0.4,trim=0 0 0 0,clip=true]{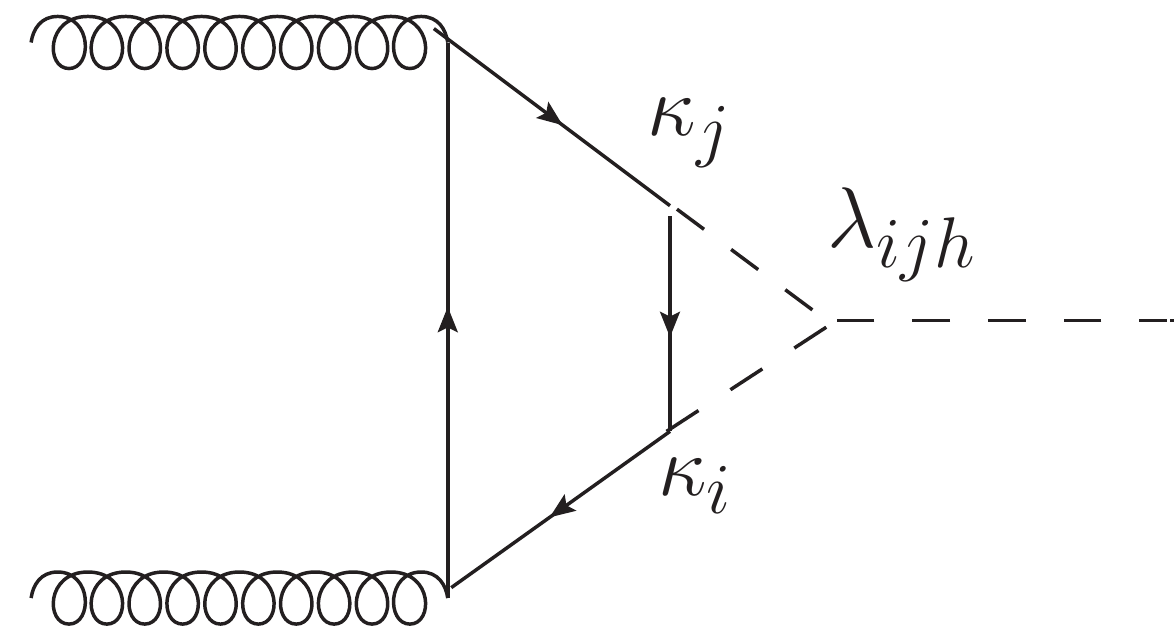}
\caption{Feynman diagrams for LO gluon fusion Higgs production (left) and NLO diagrams involving Higgs bosons (centre and right) in GxSM models.}
\label{fig:Feyn_diag_ggF}
\end{figure}
i) $A_{ggF}^{(LO)}$, the one-loop function for gluon fusion as computed for an SM-like Higgs boson; ii) $A_{iff}^{(NLO)}$, the two-loop function with a Higgs boson line inserted vertically inside the one-loop diagram connecting the two fermion lines; iii) $A_{ijf}^{(NLO)}$, the two-loop function with a fermion box, two Higgs boson lines radiating and connecting to the final state Higgs boson; and iv)  $A_{EW}^{(NLO)}$ is the two-loop function containing all the pure SM electroweak NLO corrections.
Dropping out terms that are multiplied by the suppression factors of the non-SM like Higgs and keeping only $\kappa_h$ we obtain a simplified expression
\begin{equation}
  A_{ggF}^{(NLO)}\simeq \varepsilon \kappa_h\sqrt{Z_{h}}\left[A_{ggF}^{(LO)}+\varepsilon\left(A_{hff}^{(NLO)}+\kappa_h\lambda_{hhh}A_{hhf}^{(NLO)}+A_{EW}^{(NLO)}\right)\right] \; .
\end{equation}
Squaring the amplitude and replacing $\kappa_h^2=1$ (we are keeping here terms linear in $\kappa_h$ because the sign can be important depending on the convention) we obtain
\begin{equation}
  \sigma_{ggF}^{(NLO)}\simeq \left|Z_{h}\right|\left(\sigma_{ggF}^{(LO)}+\sigma_{hff}^{(NLO)}+\kappa_h\lambda_{hhh}\sigma_{hhf}^{(NLO)}+\sigma_{EW}^{(NLO)}\right)\; ,
\end{equation}
where the $\sigma_{ggF}^{(LO)}$ is the leading order cross-section as computed in the SM and all the other $\sigma_{X}^{(NLO)}$ come from the interference between the LO amplitude, $A_{ggF}^{(LO)}$ , with the corresponding $A_{X}^{(NLO)}$ amplitude.
Re-arranging terms , defining the ratios
\begin{equation}
C_X \equiv \dfrac{\sigma_{X,{\rm SM}}^{(NLO)}}{\sigma_{ggF,{\rm SM}}^{(LO)}}
\end{equation}
and defining $|Z_h|=1+\delta Z_h$ we obtain
\begin{equation}
  \sigma_{ggF}^{(NLO)}= \sigma_{ggF}^{(LO)}\left(1+\delta_{\rm SM}+\delta_{\rm GxSM}\right)
\end{equation}
where
\begin{eqnarray}
  \delta_{\rm SM}&=&C_{hff}+C_{hhf}+C_{EW}+\delta Z_{h}^{SM}
  \\\delta_{\rm GxSM} &\simeq& \left(\frac{\kappa_h\lambda_{hhh}}{\lambda_{hhh}^{SM}}-1\right)C_{hhf}+\delta Z_{h}-\delta Z_{h}^{SM} \; . \label{eq:approx_ggF}
\end{eqnarray}
Here we denote SM limit quantities with the superscript SM so that we can separate out $\delta_{\rm SM}$, which are the NLO correction as in the SM, and the new contributions due to the extended scalar sector $\delta_{\rm GxSM}$. The factor $C_{hhf}$ was computed\footnote{In \cite{Degrassi:2016wml} $C_{hhf}$ is denoted by $C_1^{\sigma_{ggF}}$ } in~\cite{Degrassi:2016wml,Degrassi:2004mx}, so we use the value $C_{hhf}=0.0066$, which is independent of the centre of mass energy. Note that the $\kappa_h$ dependence is important because both signs are allowed in the conventions used in our scans.

We are interested in observing if there are scenarios where the new scalar contributions can correct considerably the cross-section. It is well known~\cite{Actis:2008ug} that the SM corrections are small, $\delta_{\rm SM}\simeq 5\%$. Our aim is to compute the new factor $\delta_{\rm GxSM}$. Finally, note that the SM-like parts in $\delta Z_{h}-\delta Z_{h}^{SM}$, for $T\neq S$ cancel out exactly  in the difference, in the limit we are using, so we only have to evaluate the difference over the scalar contributions without Goldstones $\left[\delta Z_{h}-\delta Z_{h}^{SM}\right]_{\rm scalars}$ using Eq.~\eqref{eq:ZfactorsGxSM}. One can check that, at one loop, $\delta Z_h=\Re(1-Z_h^{-1})$, which we can obtain from Eq.~\eqref{eq:Zqminus1}. In particular we get that in the SM limit the corresponding contribution is
\begin{equation}
\delta Z_{h}^{SM}=-\frac{9 m_h^2}{2v^2}\left(\frac{2\pi}{3\sqrt{3}}-1\right)
\end{equation}
in agreement with~\cite{Degrassi:2016wml}.

Another important issue relates to the crossing of thresholds. In particular when a threshold for the Higgs boson to decay to a pair of a lighter bosons is crossed the wave function renormalisation contains a singularity at the threshold if evaluated with a real pole mass~\cite{Passarino:2007fp,Actis:2008uh}. This problem can be cured by working with the full complex pole mass where the width of the Higgs boson in included in the imaginary part of $p^2$~\cite{Passarino:2007fp}. In our calculations -- see e.g.~Eq.\eqref{eq:Z_pert_expand} -- we expand around the real tree level masses. Thus, to avoid these unphysically large deviations, we will ignore scenarios close to these thresholds, within a $5$~GeV mass window. Away from these thresholds this approximation is known to work well -- see for example Figs.~6 and~7 of~\cite{Actis:2008uh} for the EW corrections to gluon fusion around the $WW$ threshold in the SM. The near threshold cases have the potential to produce extra enhancements. This limit is beyond the scope of our study and will be left for future work.  

In our numerical analysis we  use the following values for the relevant SM parameters (consistently with the code \texttt{sHDECAY} used to generate the tree level samples~\cite{Costa:2015llh}):
\begin{eqnarray}
m_Z&=& 91.153\; \textrm{GeV}\nonumber\\
m_W&=& 80.358 \;\textrm{GeV}\nonumber\\
y_t &=& 0.97192\nonumber\\
v &=&246.22\; \textrm{GeV} \; .
\end{eqnarray}
Note also that, in our normalisation, the SM-like Higgs triple coupling is given by
\begin{equation}
\lambda_{hhh}^{\rm SM} = 3\dfrac{m_h^2}{v} \; .
\end{equation}

\subsection{Particular models}\label{subsec:particular_models}

In Sect.~\ref{sec:Results} we analyse samples for two benchmark models to illustrate the size of the EW NLO corrections. Here we provide a brief summary of the models.
\subsubsection{The real singlet model (RxSM)}
 The potential for the model is
\begin{equation}
V_{\rm RxSM}=\dfrac{m^2}{2}H^\dagger H+\dfrac{\lambda}{4}(H^\dagger
H)^2+\dfrac{\lambda_{HS}}{2}H^\dagger H S^2 +\dfrac{m^2_S}{2} S^2 + 
\dfrac{\lambda_S}{4!}S^4 \, ,  \label{eq:V_RxSM}
\end{equation}
Here the (real) couplings of the theory are $m,\lambda,\lambda_{HS},m_S$ and $\lambda_S$ and $S$ is a real singlet field with a $\mathbb{Z}_2$  symmetry ($S\rightarrow -S$). In this model $S$ is a dark matter candidate if $v_S\equiv\left<S\right>=0$, or it is a new scalar mixing with the Higgs if $v_S\neq 0$. We will focus on the latter because the former seems to be very close
to being ruled out except in the region around $m_{h_{125}}/2$ and for very large dark matter masses 
(see for instance~\cite{Escudero:2016gzx, Wu:2016mbe, Casas:2017jjg}). The model has five independent input parameters which we choose to be $\{\alpha,m_1,m_2,v,v_S\}$ both at tree level and at one loop. Here $m_1<m_2$ are, respectively, the masses of the scalar eigenstates $h_1$ and $h_2$ decomposed as
\begin{equation}
  \left(\begin{array}{c} h_1\\h_2\end{array}\right) = \left(\begin{array}{cc}
\cos\alpha & \sin\alpha\\
-\sin\alpha & \cos\alpha
\end{array}\right)\left(\begin{array}{c}
h\\
s
\end{array}\right)+\varepsilon \left(\begin{array}{cc}
E_{\phantom{(1)2}1}^{(1)2}\left[-\sin\alpha \right. & \left.\cos\alpha\right]
\\
E_{\phantom{(1)1}2}^{(1)1}\left[\phantom{-}\cos\alpha \right. & \left.\sin\alpha\,\right]
\end{array}\right)\left(\begin{array}{c}
h\\
s
\end{array}\right)+\ldots
\end{equation}
where $\alpha \in [-\pi/2,\pi/2]$ and where we indicate the form of the the one loop correction (second term). This follows from the definitions in Eqs.~\eqref{eq:GenExpansionParameters} and~\eqref{Eq:choice mixing_offdiag}. Furthermore, one can check that the one loop system, Eq.~\eqref{eq:PoleEqs1loop}, in this model allows us to make the choice that one of the mass eigenstates is not corrected, i.e. it is input. Thus, we either set $E_{\phantom{(1)2}1}^{(1)2}$ or $E_{\phantom{(1)2}1}^{(1)2}$ to zero, respectively when $h_1=h_{125}$ or when $h_2=h_{125}$. In this way, we guarantee that there is an SM-like Higgs boson in our one-loop corrected samples that is compatible with the observed Higgs boson.  In sum, the one loop shifted parameters are then all the Lagrangian parameters, $\{m^2,\lambda,\lambda_{HS},m^2_S,\lambda_S\}$, and the eigenstate $h_i\neq h_{125}$. The VEVs were also chosen to be input, so their shift is set to zero.

\subsubsection{The complex singlet model (CxSM)}
 The potential for the model is
\begin{eqnarray}
V_{\rm CxSM}&=&\dfrac{m^2}{2}H^\dagger H+\dfrac{\lambda}{4}(H^\dagger H)^2+\dfrac{\delta_2}{2}H^\dagger H |\mathbb{S}|^2+\dfrac{b_2}{2}|\mathbb{S}|^2+
\dfrac{d_2}{4}|\mathbb{S}|^4+\nonumber\\&&+\left(\dfrac{b_1}{4}\mathbb{S}^2+a_1\mathbb{S}+c.c.\right) .  \label{eq:V_CxSM} 
\end{eqnarray}
Here $\mathbb{S}=(S+iA)/\sqrt{2}$ with $S,A$ real fields. This model has a dark phase if $v_A\equiv \left<A\right>=0$ and $v_S=0$, with $A$ the dark matter candidate and two Higgs bosons mixing. In this phase the $A\rightarrow -A$ symmetry is preserved. In the broken phase $v_A\neq 0$ and we have three Higgs bosons which may be visible at colliders. We will investigate both phases of this model.
\paragraph{Dark phase:}
In the dark phase, $v_A=0$, the set of inputs is similar to the RxSM. We choose $\{\alpha,v,v_S,a_1,m_1,m_2, m_D\}$ where $m_1<m_2$ are, respectively, the masses of the scalar eigenstates $h_1$ and $h_2$, and $m_D$ is the mass of the dark matter particle. The mass eigenstates are now decomposed as
\begin{equation}
  \left(\begin{array}{c} h_1\\h_2 \\h_D\end{array}\right) = \left(\begin{array}{ccc}
\cos\alpha & \sin\alpha & 0\\
-\sin\alpha & \cos\alpha & 0\\
0 & 0 & 1\\
\end{array}\right)\left(\begin{array}{c}
h\\
s \\
A
\end{array}\right)+\varepsilon \left(\begin{array}{ccc}
E_{\phantom{(1)2}1}^{(1)2}\left[-\sin\alpha \right. & \cos\alpha &\left.0\right]\\
E_{\phantom{(1)1}2}^{(1)1}\left[\phantom{-}\cos\alpha \right. & \sin\alpha\,& \left.0\right]  \\
\phantom{E_{\phantom{(1)1}2}^{(1)1}\left[-\right.}0 & 0 & 0
\end{array}\right)\left(\begin{array}{c}
h\\
s \\
A
\end{array}\right)+\ldots
\end{equation}
where we note that, again, we can choose to shift only the state that is not the SM like Higgs boson. After analysing the one loop system, Eq.~\eqref{eq:PoleEqs1loop}, one concludes that the set of shifted parameters can be given again by all the Lagrangian parameters except $a_1$, i.e. $\{m^2,\lambda,\delta_2,b_2,d_2,b_1\}$, and the eigenstate $h_i\neq h_{125}$. The VEVs were also chosen to be input, so their shift is zero.

\paragraph{Broken phase:}
Regarding the broken phase of the model, when $v_A\neq 0$, the input parameters are now chosen to be $\{\alpha_1,\alpha_2,\alpha_3,v,v_S,m_1,m_3\}$. The three mass eigenstates $h_1,h_2,h_3$ and their masses $m_1<m_2<m_3$ are decomposed as
\begin{equation}
  \left(\begin{array}{c} h_1\\h_2 \\h_3\end{array}\right) =
\left(\begin{array}{ccc}
c_1 c_2 & s_1 c_2 & s_2\\
-(c_1 s_2 s_3 + s_1 c_3) & c_1 c_3 - s_1 s_2 s_3  & c_2 s_3\\
- c_1 s_2 c_3 + s_1 s_3 & -(c_1 s_3 + s_1 s_2 c_3) & c_2 c_3
\end{array}\right)\left(\begin{array}{c}
h\\
s \\
a
\end{array}\right)+\ldots
\end{equation}
where $\alpha_i \in [-\pi/2,\pi/2]$ and now we do not display explicitly the choice of one loop corrections to the mass eigenstates for brevity. In this case the one loop system, Eq.~\eqref{eq:PoleEqs1loop}, forces us to introduce shifts for more than one mass eigenstate. Nevertheless, it still allows us to fix one of the mass eigenstate so, again, we choose to keep the SM-like Higgs mass and eigenstate as input. In addition we choose three corrections to the other mass eigenstates to be non-zero, such that we can solve Eq.~\eqref{eq:one_loop_mix_anti_sym} directly. For example, if the Higgs boson is $h_1$ and the other two Higgs bosons are $h_2$ and $h_3$, we can choose $E^{(1)}_{i1}=0$, and $E^{(1)}_{23}=0$ and the non-zero one loop shifts will be $E^{(1)}_{13},E^{(1)}_{12},E^{(1)}_{32}$. Finally, all the Lagrangian parameters,  $\{m^2,\lambda,\delta_2,b_2,d_2,b_1, a_1\}$, are shifted and the VEVs are not (they are again chosen to be input). 

Both the RxSM and the CxSM  were recently analysed in light of the LHC run-1 data, and compared with the NMSSM to determine if they could be distinguished from the latter in the Higgs sector~\cite{Costa:2015llh,Costa:2014qga}. We will use the samples that were generated in~\cite{Costa:2015llh}, which are compatible with all the latest theoretical and phenomenological constraints. For the case with dark matter we have also applied the latest dark matter direct detection bounds from the LUX experiment~\cite{Akerib:2016vxi}. For further details on the constraints applied in the samples we refer the reader to~\cite{Costa:2015llh}.

\subsection{Numerical results}\label{sec:Results}

In this section we use the specific models presented in the previous section to illustrate the importance of the NLO EW corrections for the the parameters of the theory and for the gluon fusion production cross-section of the observed $125$ GeV SM-like Higgs boson. In  Sect.~\ref{sec:Introduction} we have noted that, according to previous studies~\cite{Kanemura:2015fra, Bojarski:2015kra}, the NLO corrections to decay widths in the RxSM model are small, typically in the order of a few percent. Here we go beyond the simplest scalar singlet extension to compare the various models and,  at a phenomenological level, focus on the production rather than the decay.

\begin{figure}[ht!]
\centering
\mbox{\hspace{3mm}\includegraphics[scale=0.5,trim= 15 30 0 0,clip]{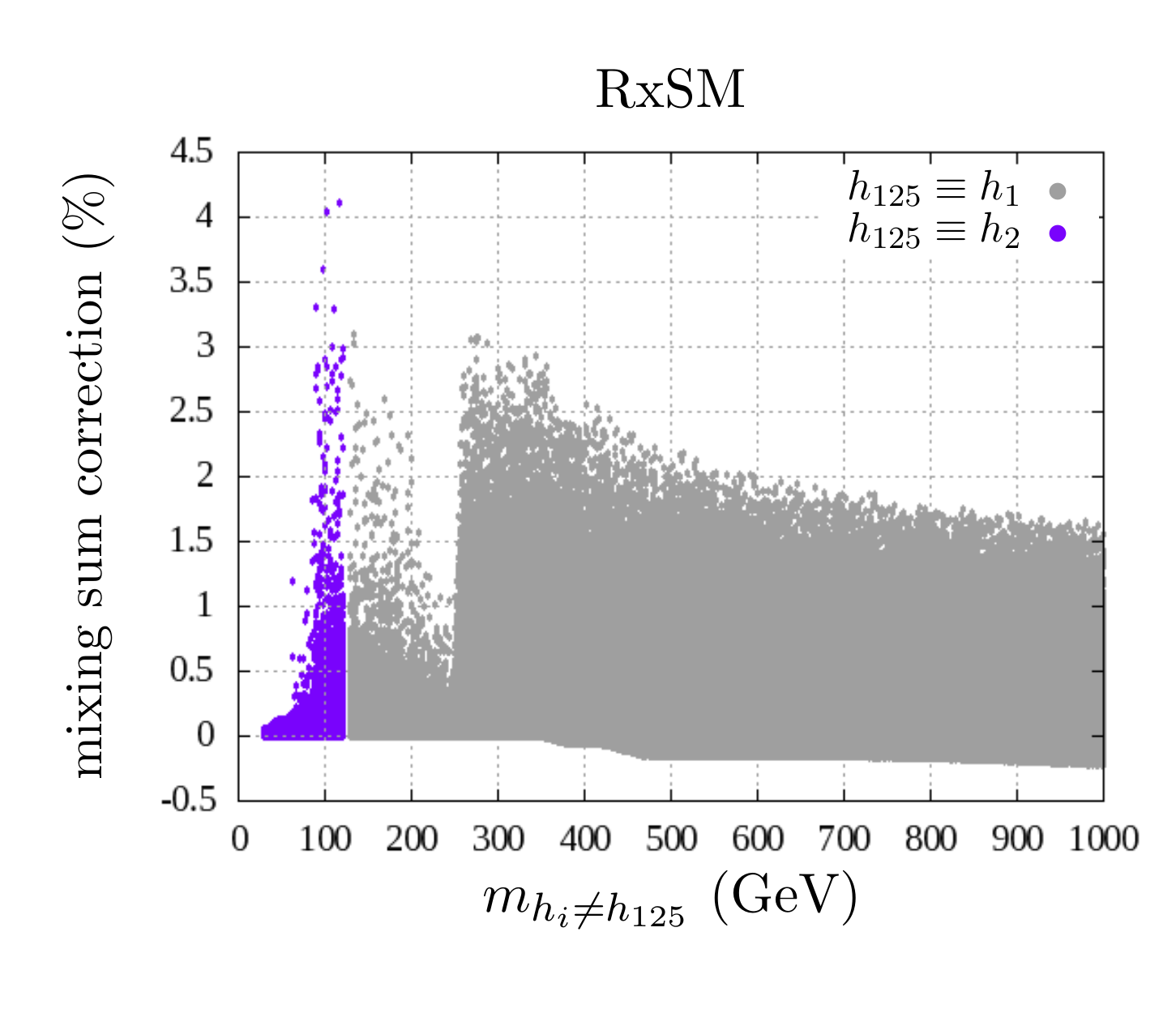}\includegraphics[scale=0.53,trim= 0 25 0 0,clip]{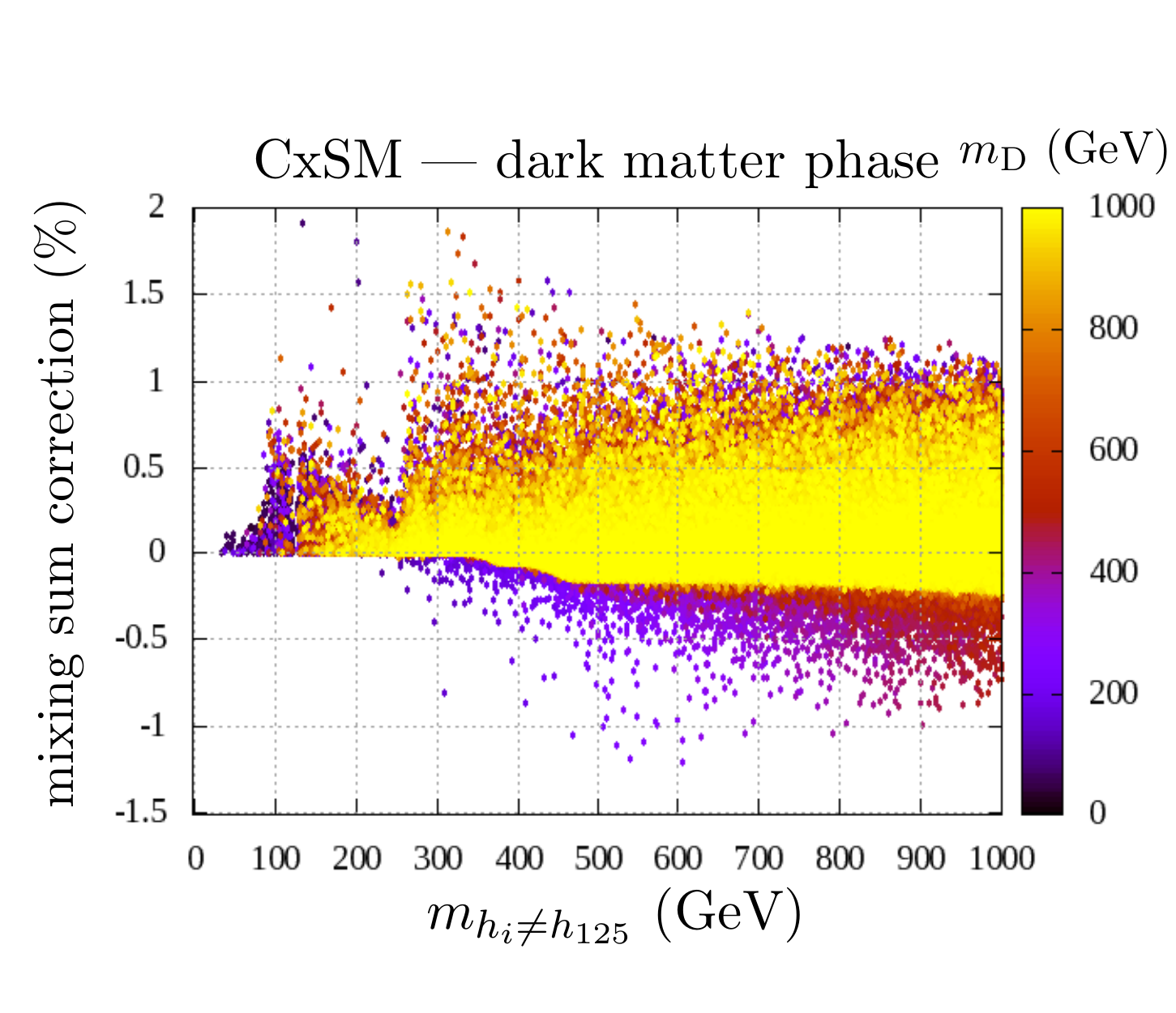}}
\mbox{\hspace{-5mm}\includegraphics[scale=0.505,trim= 0 30 0 0,clip]{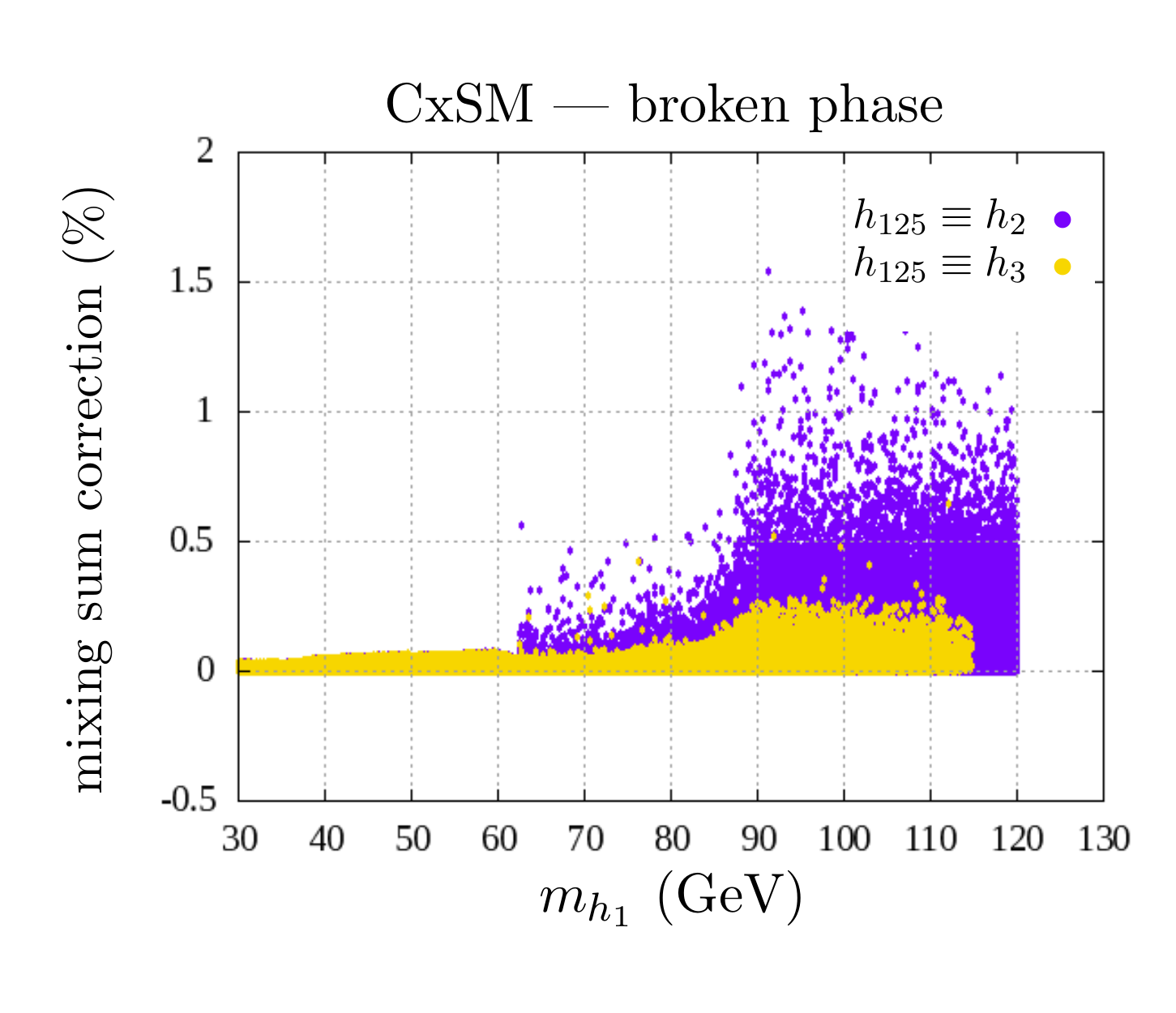}\includegraphics[scale=0.505,trim= 12 30 0 0,clip]{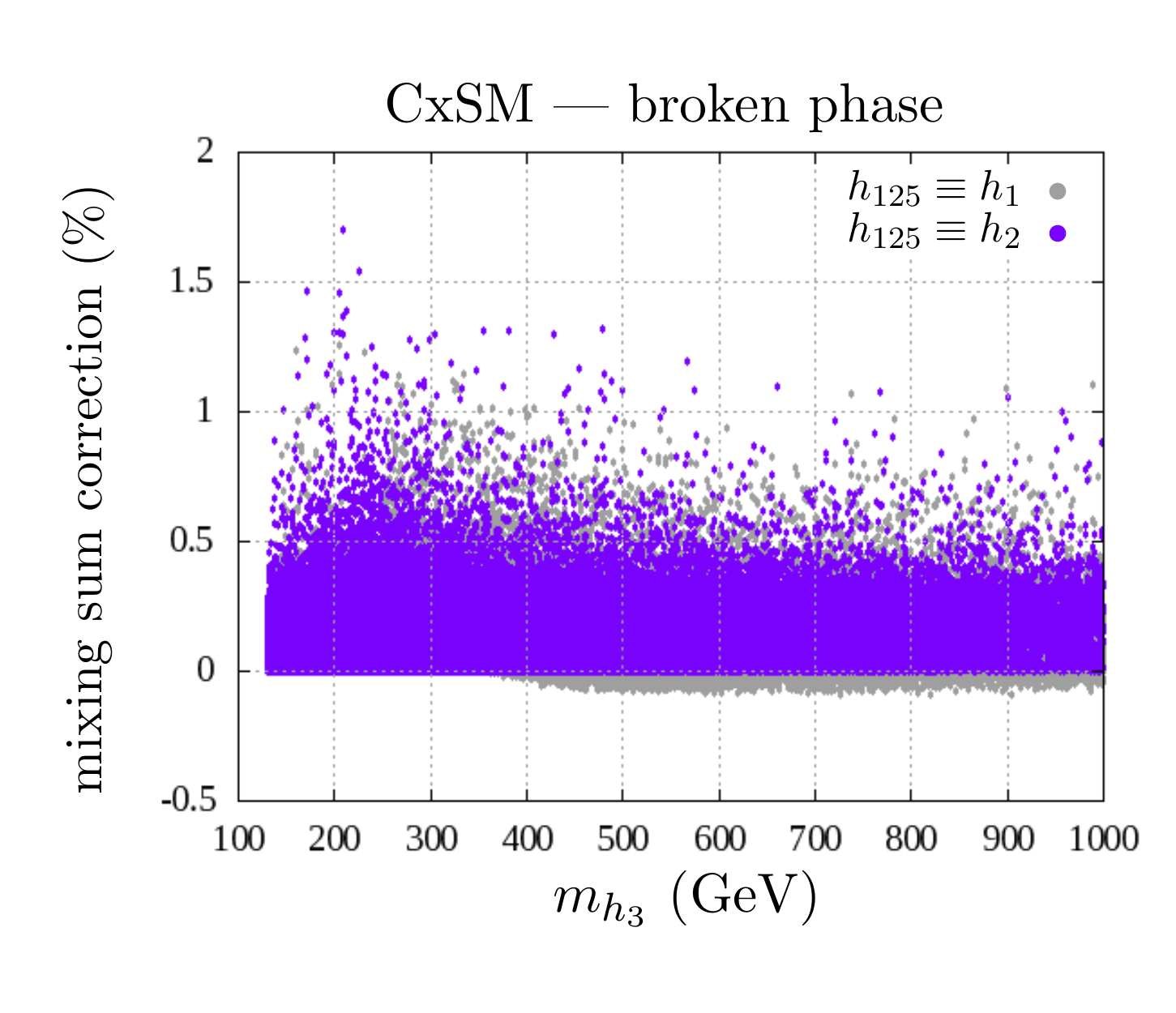}}
\caption{{\em Mixing sum corrections:} In all panels we present the one-loop mixing sum correction (vertical axis), versus the mass of one of the new non-SM like Higgs bosons (horizontal axis), for three models. {\em Top left}, the RxSM; {\em Top right}, the CxSM dark phase with the non-SM like Higgs boson mass in the horizontal axis and the dark matter mass in the colour scale; {\em Bottom}, the CxSM broken phase with either the lighter (left) or the heavier (right) of the non-SM like Higgs masses in the horizontal axis.}
\label{fig:mixsum}
\end{figure}
In Fig.~\ref{fig:mixsum} we first present the corrections to the tree level mixing sum relation. We have observed, Eq.~\eqref{eq:sum_rule_violation}, that the correction to this tree level relation is, in fact, a good proxy to evaluate the importance of the NLO corrections to the input parameters because it does not depend on further choices specific to the scheme (other than the normalisation condition of the mass eigenstates).

We start by discussing the RxSM model (top left panel). The simplicity of this model allows us to interpret the distribution of points in the scan more straightforwardly. It is also useful to interpret the distribution of points for the other models because it is a limiting case of those models. We separate the two scenarios where the SM-like Higgs is the lighter (grey) and the heavier (purple) of the two mixing Higgs bosons.  We observe that the magnitude of the correction to the tree level value of 1 for the mixing sum, ranges approximately between $+4\%$ and $-0.5\%$. The various features of the distribution of points are due to the combination of theoretical and phenomenological constraints. One can see that for small masses the mixing sum correction is small and rises sharply from $\sim m_{h_{125}}/2$~GeV to $\sim 100$~GeV as the decay channel $h_{125}\to h_1 +h_1$ closes. This is due to stronger constraints from negative searches at colliders which force the model to be more SM-like\footnote{This can also  be observed in the coloured layers of Fig.5 left of~\cite{Costa:2014qga} where the mixing Higgs spectrum and open decay channels are the same.} so that the new scalar is more decoupled (hence it also contibutes less in the loops). For masses larger than $ \sim 125$~GeV, the collider constraints become progressively more restrictive up to the opening of the decay of the heavy Higgs to a pair of SM-like Higgs bosons ($h_{125}$) where there is again a sharp rise. The top boundary for large masses is due to the electroweak precision data constraints. For the lower boundary, we observe that negative corrections become possible at about $\sim 350$~GeV, i.e.  the threshold for decay to a pair of top quarks.

For the CxSM dark phase (right panel), we observe a distribution of points that is similar to the RxSM with the difference that larger magnitude negative values are allowed. The colour scale, which represents the dark matter mass, shows that this is possible in scenarios where the dark matter particle running in the loops is lighter than about $\sim 500$~GeV. Observing the yellow points, corresponding to large dark matter masses, we recover the lower boundary observed for the RxSM (see yellow region). This is expected because the contributions from the dark matter loop propagators are suppressed for large masses.

Regarding the CxSM broken phase (bottom panels) the distribution of points is more complicated because we have three mixing Higgs bosons. Nevertheless, from the vertical axes, we see that the magnitude of the upper and lower ranges of the mixing sum correction is similar. In the two panels we indicate the scenario where the SM-like Higgs is the lightest, next to lightest and heaviest of the three mixing Higgs bosons, respectively with grey, purple and yellow points. In the left panel we have the smallest mass in the horizontal axis and in the right panel the largest mass. In the two scenarios of the left panel (yellow and purple points) we observe that the points distribute similarly to the RxSM. This is consistent with the observation, in the RxSM, that a light scalar in the loop corrections produces positive corrections. The only difference is that the yellow points in the peak region for masses larger than $\sim 62$~GeV are more suppressed. This is consistent with the fact that the SM coupling is diluted over two small mixing factors for each of the two light scalars, which further suppresses each contribution. The yellow points stop at $\sim 118$~GeV because we have applied a cut to avoid degenerate scenarios with a distance of $3.5$~GeV between any two scalar states. This is why the purple layer ($h_{125}\equiv h_2$) stops at $\sim 121$~GeV in the left panel.
In the right panel we can observe the scenario $h_{125}\equiv h_1$ in the grey points. There we see that for scenarios where $h_3$ is heavier than about $350$~GeV we can have a negative correction. This is consistent with the RxSM observation that negative corrections are possible in this scenario when the heavy scalar is above this mass. The main difference with the RxSM is that the lower boundary is not so sharply defined. This is simply due to loss in density in the scan for the CxSM, which has more parameters making it harder to collect large amounts of points. Finally, note that for the purple points, where the SM-like Higgs is the next to lightest, it is not possible to obtain a negative correction, which indicates that the positive contributions from one lighter Higgs are enough to push the correction to positive values. Also note that, contrarily to the RxSM plot, we cannot see any sharp rise at $m_{h_3}\sim 2 m_{h_{125}}$ because it is always possible, for fixed $m_{h_3}$, to have decays involving a lighter non-SM like Higgs with a different mass. This extra free parameter in the scan dilutes such a boundary.

Now we turn to the discussion of the NLO EW scalar contributions to the corrections to gluon fusion in the limit, already discussed in Sect.~\ref{sec:gluon_fusion_xs}, where the mixing factor of the SM-like Higgs boson is close to the SM limit $\kappa_{h_{125}}^2 \to 1$. In our numerical analysis we have evaluated, for each model, the quantity $\delta_{\rm GxSM}$, in Eq.~\eqref{eq:approx_ggF} specialised to the RxSM and CxSM models using the samples already discussed. In addition, we have selected points within $10\%$, $5\%$ and $2\%$ of the limiting case $\kappa_{h_{125}}^2\to 1$. Close to this limit our approximation is then reliable. Furthermore, applying this increasingly tighter constraint simulates the experimental situation where the SM-like Higgs boson couplings are measured to be SM-like with an increasingly higher accuracy. Understanding how large the new scalar corrections are allowed to be, when such tight experimental constraints become available, may then provide improved bounds on each model. Finally, in all plots, we have applied a $5$~GeV mass window exclusion around thresholds for the opening of SM-like Higgs to scalars decays to avoid the singular behaviour of the wave function renormalisation discussed at the end of Sect.~\ref{sec:gluon_fusion_xs}. 

\begin{figure}[t]
\centering
\includegraphics[scale=.54,trim= 15 30 0 0,clip]{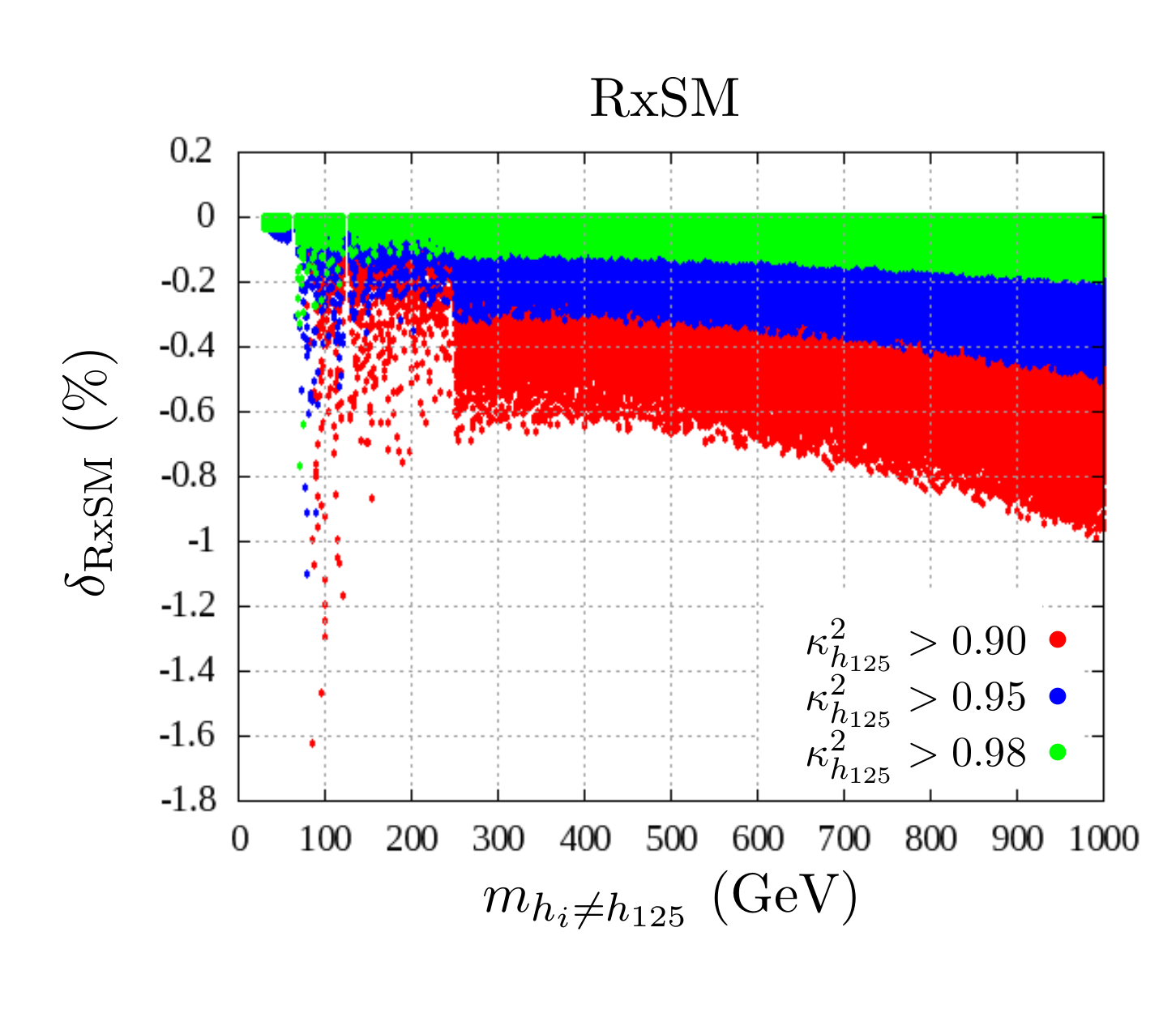}\includegraphics[scale=.54,trim= 15 30 0 0,clip]{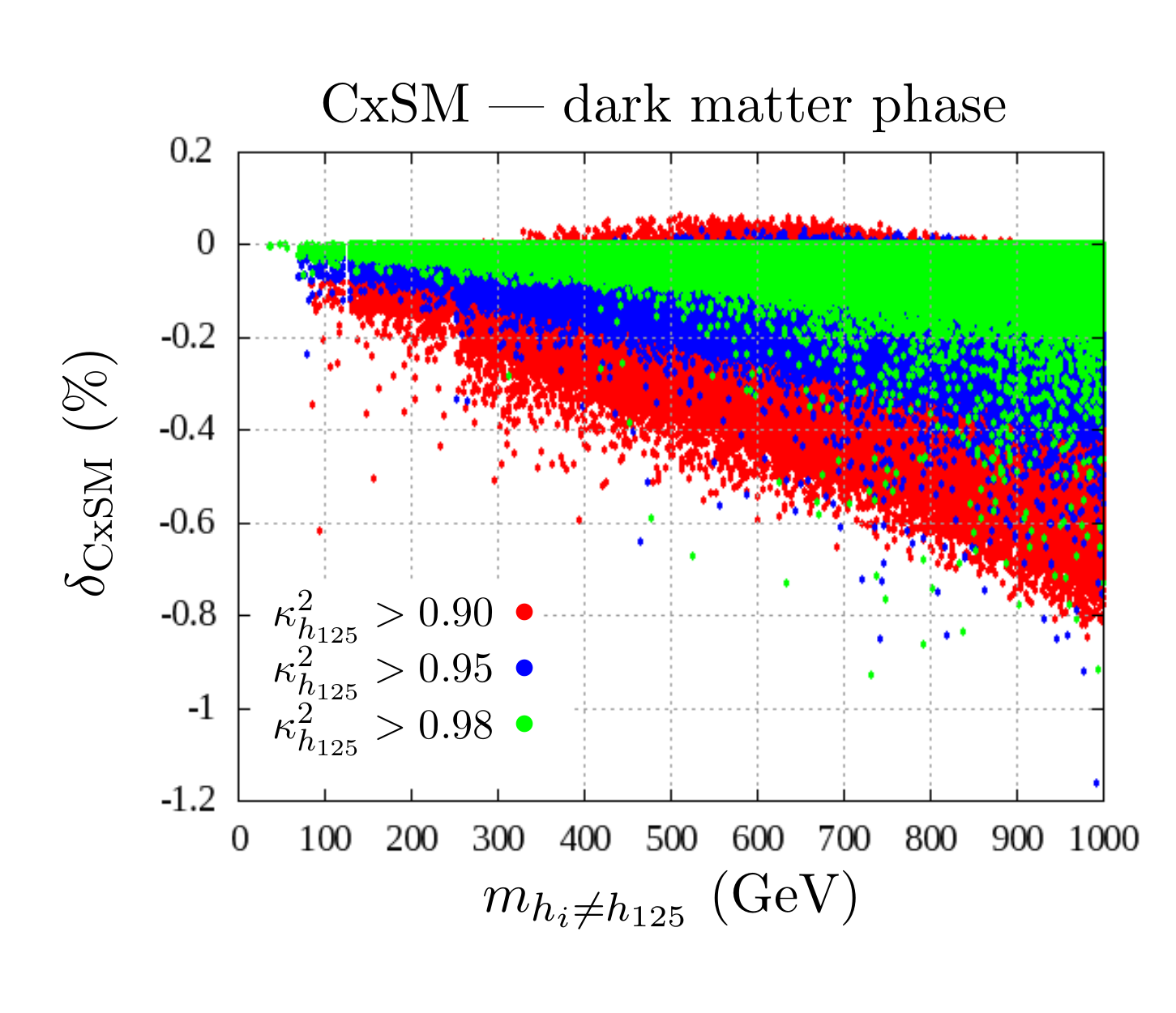}
\includegraphics[scale=.54,trim= 15 30 0 0,clip]{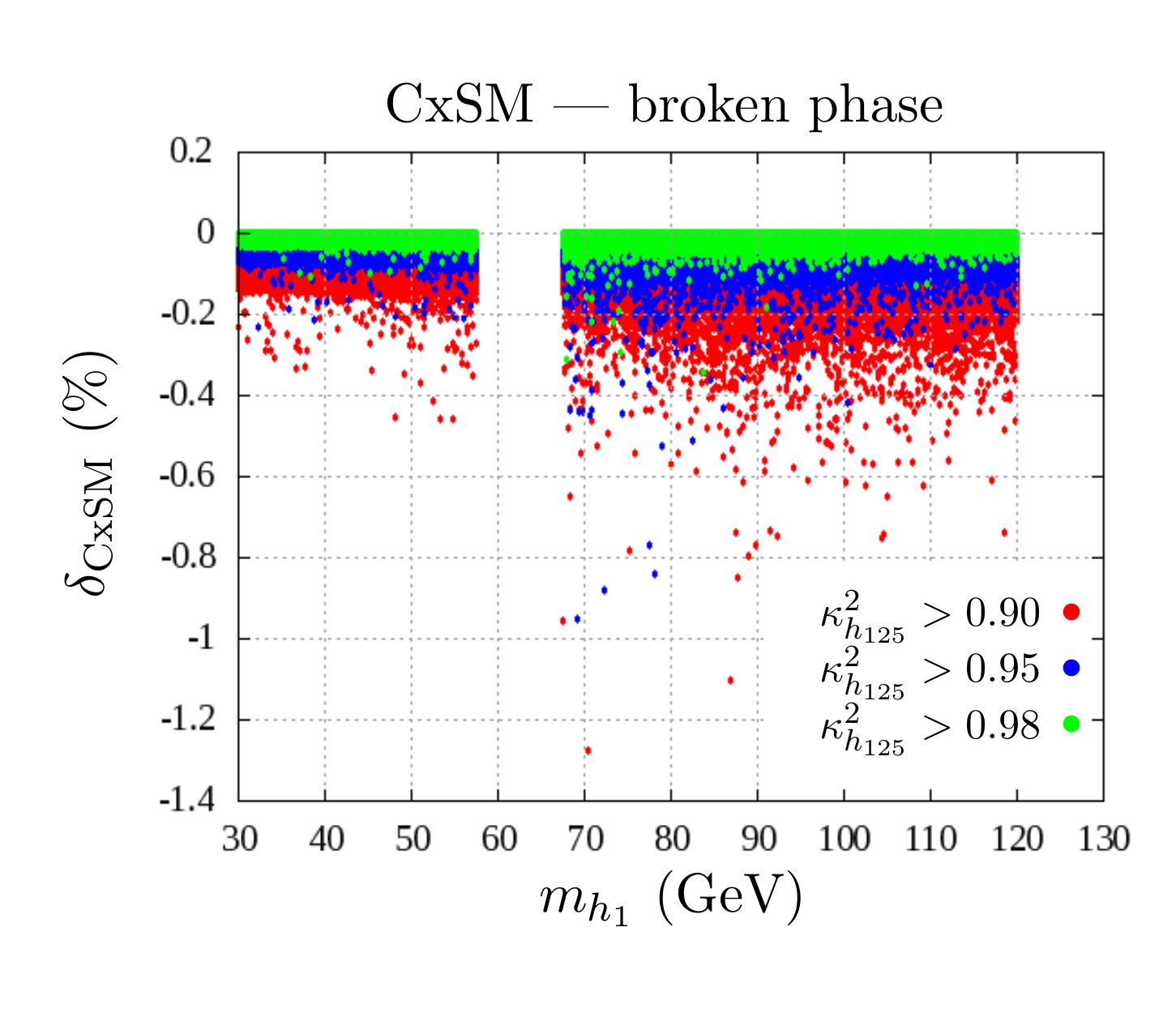}\includegraphics[scale=.54,trim= 15 30 0 0,clip]{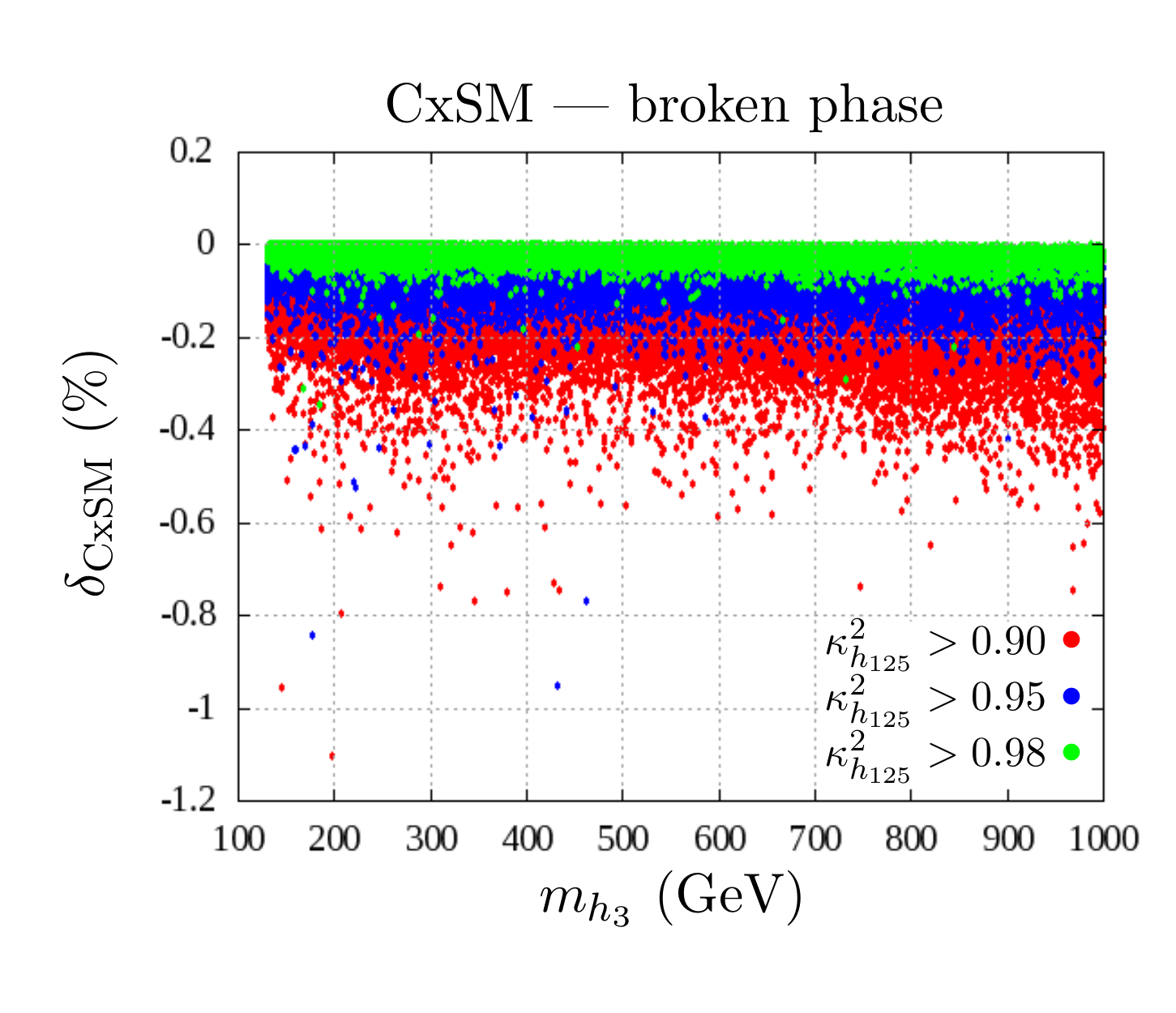}
\caption{{\em Gluon fusion corrections:} In all panels we present the NLO EW scalar contribution to the correction to gluon fusion production of the SM-like Higgs (vertical axis), versus the mass of one of the new non-SM like Higgs bosons (horizontal axis), for the three models. {\em Top left}, the RxSM; {\em Top right}, the CxSM dark phase with the non-SM like Higgs boson mass in the horizontal axis; {\em Bottom}, the CxSM broken phase with either the lighter (left) or the heavier (right) of the non-SM like Higgs masses in the horizontal axis.}
\label{fig:deltaGxSM}
\end{figure}

In Fig.~\ref{fig:deltaGxSM} we can first observe, by inspecting the vertical axes for each model, that the NLO EW scalar contributions to the corrections for the two models without dark matter (top left and bottom panels) are always negative, whereas for the CxSM dark (top right) positive values are possible as we move away from the limit $\kappa^2_{h_{125}}\to 1$ (blue and red points). Focusing first on the RxSM (top left) we observe that the distribution of points away from the SM-limit and the various peaks and thresholds follow closely the same patterns already observed in Fig.~\ref{fig:mixsum}. For masses of the new Higgs in the range $\sim 70$~GeV to $\sim 100$~GeV we observe that the corrections can deviate away from the SM by about $-1.6\%$, $-1.1\%$ and $-0.8\%$ for the red, blue and green points respectively.  For larger masses, above the SM-like Higgs mass, we observe well defined boundaries for the three layers and the correction can be at most about $-1\%$, $-0.5\%$ and $-0.2\%$, respectively for the red blue and green points. For the broken phase of the CxSM (bottom panels), the ranges of values for the corrections are similar. In the bottom left panel, with the lightest Higgs boson mass on the horizontal axis, we can see more clearly the exclusion mass window we have applied around $m_{h_{125}}/2$ to avoid dealing with the singular limit. In the bottom right panel we represent the same points as a function of the largest mass.  By observing the two bottom panels we do not recover directly the light and the heavy region of the RxSM plot since the points spread down not only for small values of the masses. This is because we have more than one Higgs boson coexisting with the SM-like Higgs.
For the bottom right panel, we can have one scenario where $h_{1}\equiv h_{125}$ and another one where $h_{2}\equiv h_{125}$. In the latter scenario we have checked that we can have $h_1$ in the mass region around $100$~GeV to contribute to the correction with a larger negative value in a way similar to the RxSM, whereas the other scenario is very similar to the RxSM. This mixture of scenarios and the fact that the CxSM has more free parameters explains the distribution of points and the absence of sharply defined boundaries in the scatter plots.

Finally, we discuss the dark phase of the CxSM which is the one that allows for larger deviations even for the scenarios closest to the SM-like limit. In the top right panel of Fig.~\ref{fig:deltaGxSM} we see that the green points can spread down to negative values as negative as the other two layers. Furthermore there are also points where the correction can be positive. We have checked that these deviation, even in the limit very close to the SM, are possible due to contributions from light to intermediate mass dark matter in the region $m_D\lesssim 500$~GeV similarly to Fig.~\ref{fig:mixsum} (top panel). For heavier dark matter scenarios we recover the RxSM distribution of points.

\section{Conclusions}
\label{sec:conclusions}

In this work we have developed a general framework for the calculation of NLO EW corrections in scalar extensions
of the SM with focus on models with an arbitrary number of scalar singlet fields added to the SM field content. The derived  set of equations can be applied for a wide class of choices of renormalisation scheme with the only restriction that the chosen input parameters consistently allow for a solution of the truncated linear system -- see also Eq.~\eqref{eq:PoleEqs1loop} and related discussion. Our results go beyond the effective potential approach since they are valid for any external momentum in the inverse propagators. We also point out that our general procedure is well suited for the automation of such corrections within a general purpose numerical tool such as \texttt{ScannerS}~\cite{ScannerS,Coimbra:2013qq}.

We then applied our method to specific models: the real scalar singlet extension and the complex scalar singlet extension of the SM. First, in order to assess the importance of the NLO EW corrections to the parameters of the theories, we calculated the NLO correction to the mixing sum $\sum_{x}\kappa_x^2$, which is one at tree-level. Other than the normalisation condition for the one-loop mass eigenstates this quantity is scheme independent. Thus it provides a good measure of the 
general trend for the magnitude of the one-loop corrections (in this case to a tree level mixing sum relation). The correction we found for this sum was at most 4\%, already hinting that, for physical processes, the NLO EW corrections are small.

Thus, we then moved on to the evaluation of the NLO EW corrections for a physical process: gluon fusion production of the SM-like Higgs. We worked in the limit where the SM-like Higgs has couplings to SM particles close to the SM values. This is the preferred region of the allowed parameter space given the latest Higgs signal measurements from the LHC. We separated out the new contributions due to the extended scalar sector from the fixed SM-like NLO EW contributions ($\sim 5\%$). We found that, similarly to earlier results for the decays in the real singlet model, the NLO EW corrections to gluon fusion production are of the order of a few percent in all models. In fact, even though we have examined three different scenarios, one in the RxSM and two in the CxSM, the general conclusion is similar: the new scalar sector corrections range between about $\sim -2\%$ and $\sim 0.1 \%$ (or between about $\sim 3\%$ and $\sim 5\%$ if we sum the SM-like contribution). In all scenarios without dark matter, we also observe that the more we approach the SM-like limit, the smaller the new scalar sector corrections, thus consistently recovering the SM NLO EW correction. The exception is the dark matter phase of the CxSM where the dark matter particle can produce a non-zero correction even in that limit. Nevertheless, in this model the overall effect is to shift the SM correction from $5\%$ to about $4\%$. The main conclusion of this study is then that future improvements in the measurements of gluon fusion production of the SM-like Higgs in these singlet models will, likely, not be able to probe radiative effects due to the new scalars. Combined with previous results in real singlet studies for the decay, where small corrections were also found, and with the fact that interference effects can be large, such minimal scalar singlet extensions will not be easily probed in future Higgs boson precision measurements. Thus, these minimal scalar singlet extended models have to be probed through direct searches for the new particles in their spectrum.

Despite the smallness of the corrections that we have found there are a few open questions. With a new dark scalar in the spectrum we have found that the corrections can deviate from the SM. In scenarios with multi-singlet dark matter, could the corrections be large enough to shift the Higgs boson properties visibly within the precision of future measurements? On the other hand we have not studied the mass region near the threshold where the SM-like Higgs can decay to a pair of scalars. Can we have a considerable enhancement of the corrections close to this threshold? This could be particularly interesting because the corresponding Higgs decay channel to a pair of scalars is kinematically suppressed near the threshold making it unlikely to be observed directly. But in the dark case this is precisely the region where the Higgs-dark-dark coupling is allowed to be larger by the dark matter relic density constraints\footnote{This is because a larger coupling allows for a more efficient dark matter annihilation in the early Universe through this channel, thus avoiding over-shooting the measured value from the Planck satellite data}. These, and other questions, will be left for future investigations.

\section*{Acknowledgements}
The authors thank Philipp Basler for interesting discussions on coincidence limits and IR divergences.
 M.S. is funded through the grant BPD/UI97/5528/2017. The work in this paper was also supported by the CIDMA project UID/MAT/04106/2013 and also by HARMONIA National Science Center - Poland project UMO-2015/18/M/ST2/00518.

\appendix

\section{Relations between bases}
\label{app:RelBases}
The $L$-basis relates to the $\Lambda$-basis as follows. The field dependent scalar couplings are
\begin{eqnarray}
\Lambda & \equiv & L^{i}v_{i}+\dfrac{1}{2!}L^{ij}v_{i}v_{j}+\dfrac{1}{3!}L^{ijk}v_{i}v_{j}v_{k}+\dfrac{1}{4!}L^{ijkl}v_{i}v_{j}v_{k}v_{l}=V^{(0)}(v_i)\;,\nonumber\\
\Lambda^{i}_{(S)} & \equiv & L^{i}+L^{ij}v_{j}+\dfrac{1}{2}L^{ijk}v_{j}v_{k}+\dfrac{1}{6}L^{ijkl}v_{j}v_{k}v_{l}\;,\nonumber\\
\Lambda_{(S)}^{ij} & \equiv & L^{ij}+L^{ijk}v_{k}+\dfrac{1}{2}L^{ijkl}v_{k}v_{l}\;, \\
\Lambda^{ijk}_{(S)} & \equiv & L^{ijk}+L^{ijkl}v_{l}\;,\nonumber\\
\Lambda^{ijkl}_{(S)} & \equiv & L^{ijkl} \nonumber\;,
\end{eqnarray}
 and the field dependent gauge couplings are
\begin{eqnarray}
\Lambda_{(G)}^{ab} & \equiv & \dfrac{1}{2}G^{abij}v_{i}v_{j}\;,\nonumber\\
\Lambda^{abi}_{(G)} & \equiv & G^{abij}v_{j}\;,\\
\Lambda^{abij}_{(G)} & \equiv & G^{abij} \nonumber \; .
\end{eqnarray}
 We have defined the mass-squared matrices, $\Lambda_{(S)}^{ij}$, and $\Lambda_{(G)}^{ab}$. For fermions it is a hermitian matrix
\begin{equation}\label{eq:F_MassSquared}
\Lambda_{(F)}^{IJ} \equiv  M^{*IL}M_{L}^{\phantom{L}J}
\end{equation}
with the (symmetric) fermion mass matrix 
\begin{equation}\label{eq:F_Mass}
M^{IJ}=Y^{IJ}+Y^{IJk}v_{k}\;.
\end{equation}
We also define fermionic cubic and quartic effective vertices, involving two scalars and two fermions, which are hermitian with respect to fermionic indices:
\begin{eqnarray}
 \Lambda^{IJk}_{(F)}	&\equiv&	Y^{*ILk}M_{L}^{\phantom{L}J}+M^{*IL}Y_{L}^{\phantom{L}Jk} \nonumber\\
\Lambda^{IJkm}_{(F)}	&\equiv&	Y^{*ILk}Y_{L}^{\phantom{L}Jm}+Y^{*ILm}Y_{L}^{\phantom{L}Jk} \; . \label{Eq:Lambda-couplings-derivatives}
 \end{eqnarray}
The matrices used to rotate from the $\Lambda$-basis to the $\lambda$-basis are defined through 
\begin{eqnarray}
R_{i}&=&\left[O_{(S)}\right]_{\phantom{j}i}^{j}\phi_{j}\nonumber \\
A_{a\mu}&=&\left[O_{(G)}\right]_{\phantom{b}a}^{b}\mathcal{A}_{b\mu}\label{Eq:Rot-fields}\\
\psi_{I}&=&\left[U_{(F)}^{*}\right]_{\phantom{J}I}^{J}\Psi_{J} \; . \nonumber
\end{eqnarray}

\section{Loop functions}\label{app:LoopFuncs}
Here we present a summary of the loop functions that we have used, which can be obtained from~\cite{Martin:2003qz}. The basic scalar loop function that we use is\footnote{Here the factor $i\epsilon$ is an infinitesimal quantity to define the integration contour on the complex plane.}
\begin{eqnarray}\label{eq:Bxy}
B_s(x,y)&\equiv& -\int_{0}^{1}dt\logbar\left[tx+(1-t)y-t(1-t)s-i\epsilon\right]\\
&=& 2-\logbar s+\begin{cases}
\sum_{k=\pm}\left\{\left(t_{k}-1\right)\log\left|1-t_{k}\right|-t_{k}\log\left|t_{k}\right|\right\}+i\pi\delta t & \quad,\Delta>0\vspace{3mm}\\
\sqrt{|\Delta|}\left(\arctan\left[\frac{2c}{\sqrt{|\Delta|}}\right]-\arctan\left[\frac{2\left(1+c\right)}{\sqrt{|\Delta|}}\right]\right)+&\vspace{1mm}\\
c\log\left(\dfrac{\left|\Delta\right|}{4}+c^{2}\right)-\left(1+c\right)\log\left(\dfrac{\left|\Delta\right|}{4}+\left(1+c\right)^{2}\right)& \quad,\Delta\leq0
\end{cases}\nonumber
\end{eqnarray}
where
\begin{eqnarray}
\Delta&\equiv&\frac{s^2+x^{2}+y^{2}-2(sx+sy+xy)}{s^2}\\
t_\pm&\equiv&\frac{s-x+y\pm s\sqrt{\Delta}}{2s}\\
 c&\equiv&\frac{x-y-s}{2s} \\
\delta t&\equiv&
\begin{cases}
1-t_{-} & \quad,\left(t_{+}>1\right)\wedge\left(0<t_{-}<1\right)\\
1 & \quad,\left(t_{+}>1\right)\wedge\left(t_{-}<0\right)\\
t_{+} & \quad,\left(0<t_{+}<1\right)\wedge\left(t_{-}<0\right)\\
\sqrt{\Delta} & \quad,0<t_{-}<t_{+}<1\\
0 & \quad,{\rm otherwise}
\end{cases}\; .
\end{eqnarray}
The various limits that are necessary are:
\begin{eqnarray}
B_0(x,y)&=&1-f^{(1)}(x,y)\\
B_s(x,0)=B_s(0,x)&=&2-\logbar x+\left(\frac{x}{s}-1\right)\left[\log\left|1-\frac{s}{x}\right|-i\pi\theta\left(1-\frac{x}{s}\right)\right] \\
B_s(0,0)&=&2-\logbar s+i\pi
\end{eqnarray}
The scalar loop function is
\begin{equation}
\Delta B_{SS}\left(x,y\right)=\int_{0}^{1}dt\log\left[\dfrac{tx+(1-t)y-t(1-t)s}{tx+(1-t)y}\right]=B_0\left(x,y\right)-B_s\left(x,y\right)\equiv -\Delta B_s(x,y)
\end{equation}
We can also obtain the limiting case $\epsilon\rightarrow 0$
\begin{equation}\label{Eq:Bssepsilonepsilon}
\Delta B_{SS}(\epsilon,\epsilon)\rightarrow -2+\logbar s-\logbar\epsilon -i\pi \;.
\end{equation}
The fermionic functions are
\begin{eqnarray}
\Delta B_{FF}(x,y)&=& (x+y)\Delta B_s(x,y) -s\, B_s(x,y)\\
\Delta B_{\bar{F}\bar{F}}(x,y)&=&2\Delta B_s(x,y)  
\end{eqnarray}
and the corresponding $\epsilon \rightarrow 0$ limits are
\begin{eqnarray}\label{Eq:BFFepsilonepsilon}
\Delta B_{FF}(\epsilon,\epsilon)&=& s\left(-2+\logbar s-i\pi\right)\\
\Delta B_{\bar{F}\bar{F}}(\epsilon,\epsilon)&=&-2\left(-2+\logbar s-\logbar\epsilon -i\pi \right) \; . 
\end{eqnarray}
Finally, the loop functions involving vector bosons are
\begin{eqnarray}
  \Delta B_{SV}(x,y)&=&(2x-y)\Delta B_s(x,y)+2sB_s(x,y)-\frac{s}{y}A(y)+ \\
  &&+\frac{s\left(2x-s\right)}{y}\left[B_s(x,y)-B_s(x,0)\right]-\frac{x^{2}}{y}\left(\Delta B_s(x,y)-\Delta B_s(x,0)\right) \nonumber
\end{eqnarray}
and
\begin{eqnarray}
\Delta B_{VV}(x,y)&=&-\frac{5}{2}\Delta B_s(x,y)+\frac{1}{4xy}\left[s(2x+2y-s)B_s(x,y)-(x^{2}+y^{2})\Delta B_s(x,y)+\right.\nonumber\\&&\left.-s(2x-s)B_s(x,0)+x^{2}\Delta B_s(x,0)\right]\nonumber\\&&-\frac{1}{4xy}\left[s(2y-s)B_s(0,y)-y^{2}\Delta B_s(0,y)+s^{2}B_s(0,0)\right]\;,
\end{eqnarray}
with
\begin{equation}
A(x)\equiv x\left(\logbar x-1\right)\; .
\end{equation}
One can check that $\Delta B_{SV}(0,0)$ is finite and that, as $\epsilon\rightarrow 0$, 
\begin{eqnarray}\label{Eq:BVVepsilonepsilon}
\Delta B_{VV}(\epsilon,\epsilon)&\rightarrow&-3\left(\frac{3}{2}-\log\frac{s}{\mu^{2}}+\log\frac{\epsilon}{\mu^{2}}+i\pi\right)\;.
\end{eqnarray}

Finally, the derivatives of the loop functions that are necessary to obtain the wave function renormalisation factors can all be expressed in terms of the following derivative
\begin{eqnarray}
  \partial_sB_s(x,y)&\equiv& \int_{0}^{1}dt\dfrac{t(1-t)}{tx+(1-t)y-t(1-t)s-i\epsilon} \\
  &&\nonumber\\
  &=& \begin{cases}
    -\dfrac{1}{s}+\dfrac{1}{\sqrt{\Delta}s}\sum_{k=\pm}\,t_{k}\left(1-t_{k}\right)\left[k\log\left|1-t_{k}^{-1}\right|+\pi i\theta\left(1-t_{k}\right)\theta\left(t_{k}\right)\right] \hspace{-2mm}& ,\Delta\geq0\\
    & \\
    -\dfrac{1}{s}+\dfrac{1}{s}\left(c+\frac{1}{2}\right)\log\left(\frac{\left(1+c\right)^{2}-\frac{\Delta}{4}}{c^{2}-\frac{\Delta}{4}}\right)-&\\
    -\dfrac{1}{s}\left(c(1+c)+\frac{\Delta}{4}\right)\frac{2}{\sqrt{-\Delta}}\left[\arctan\left(\frac{2(1+c)}{\sqrt{-\Delta}}\right)-\arctan\left(\frac{2c}{\sqrt{-\Delta}}\right)\right] & ,\Delta<0\; .
\end{cases}
\end{eqnarray}

\section{Some useful identities}\label{app:identities}
In this section we prove a few useful identities. We first want to relate a contraction between the effective cubic fermion-fermion-scalar couplings, with a contraction of the Yukawa couplings with the mass matrices for massless fermion states as follows:
\begin{eqnarray}
\lambda_{(F)\phantom{E_{2}}i}^{E_{1}E_{2}}\lambda_{(F)E_{2}E_{1}j}+c.c.&=&\left(y_{\phantom{E_{1}}Li}^{\star E_{1}}m^{LE_{2}}+m^{\star E_{1}L}y_{L\phantom{E_{2}}i}^{\phantom{L}E_{2}}\right)\left(y_{\phantom{\star}E_{2}\phantom{K}j}^{\star\phantom{E_{2}}K}m_{KE_{1}}+m_{\phantom{\star}E_{2}K}^{\star}y_{\phantom{K}E_{1}j}^{K}\right)+c.c.\nonumber\\
&=&y_{\phantom{E_{1}}Li}^{\star E_{1}}m^{LE_{2}}y_{\phantom{\star}E_{2}\phantom{K}j}^{\star\phantom{E_{2}}K}m_{KE_{1}}+y_{\phantom{E_{1}}Li}^{\star E_{1}}m^{LE_{2}}m_{\phantom{\star}E_{2}K}^{\star}y_{\phantom{K}E_{1}j}^{K}+\nonumber\\
&&m^{\star E_{1}L}y_{L\phantom{E_{2}}i}^{\phantom{L}E_{2}}y_{\phantom{\star}E_{2}\phantom{K}j}^{\star\phantom{E_{2}}K}m_{KE_{1}}+m^{\star E_{1}L}y_{L\phantom{E_{2}}i}^{\phantom{L}E_{2}}m_{\phantom{\star}E_{2}K}^{\star}y_{\phantom{K}E_{1}j}^{K}+c.c.\nonumber\\
&=&y_{\phantom{E_{1}}Li}^{\star E_{1}}m^{LE_{2}}y_{\phantom{\star}E_{2}\phantom{K}j}^{\star\phantom{E_{2}}K}m_{KE_{1}}+y_{\phantom{E_{1}}Li}^{\star E_{1}}m^{LE_{2}}m_{\phantom{\star}E_{2}K}^{\star}y_{\phantom{K}E_{1}j}^{K}+\nonumber\\
&&m^{E_{1}L}y_{L\phantom{E_{2}}i}^{\star E_{2}}y_{E_{2}\phantom{K}j}^{\phantom{E_{2}}K}m_{KE_{1}}^{\star}+m^{E_{1}L}y_{L\phantom{E_{2}}i}^{\star E_{2}}m_{\phantom{\star}E_{2}K}y_{\phantom{\star K}E_{1}j}^{\star K}+c.c.\nonumber\\
&=&y_{\phantom{E_{1}}Li}^{\star E_{1}}m^{LE_{2}}y_{\phantom{\star}E_{2}\phantom{K}j}^{\star\phantom{E_{2}}K}m_{KE_{1}}+y_{\phantom{E_{1}}Li}^{\star E_{1}}m^{LE_{2}}m_{\phantom{\star}E_{2}K}^{\star}y_{\phantom{K}E_{1}j}^{K}+\nonumber\\
&&m^{E_{2}L}y_{L\phantom{E_{2}}i}^{\star E_{1}}y_{E_{1}\phantom{K}j}^{\phantom{E_{2}}K}m_{KE_{2}}^{\star}+m^{E_{2}L}y_{L\phantom{E_{2}}i}^{\star E_{1}}m_{\phantom{\star}E_{1}K}y_{\phantom{\star K}E_{2}j}^{\star K}+c.c.\nonumber\\
&=&2y_{\phantom{E_{1}}Li}^{\star E_{1}}m^{LE_{2}}y_{\phantom{\star}E_{2}\phantom{K}j}^{\star\phantom{E_{2}}K}m_{KE_{1}}+2y_{\phantom{E_{1}}Li}^{\star E_{1}}m^{LE_{2}}m_{\phantom{\star}E_{2}K}^{\star}y_{\phantom{K}E_{1}j}^{K}+c.c.\; .\nonumber
\end{eqnarray}
In the third equality we have use the fact that the c.c. allows us to complex conjugate its second line (which is equivalent to swapping with the terms hidden in the c.c.). In the fourth equality we have relabelled the dummy indices $E_{1}\rightarrow E_{2}$ and $E_{2}\rightarrow E_{1}$. In the last line we have used the symmetry of of the involved tensors under exchange of the fermionic indices . Now noting that, in the mass-squared eigenbasis, the $m_{AB}$ can be nonzero only if $m_{A}=m_{B}$\footnote{This is a consequence of the fact that $\Lambda_{(F)}=M\,M^\dagger$ (using matrix notation both for $\Lambda_{(F)AB}$ and $M_{AB}$), so the invariant subspace associated with each eigenvalue of $\Lambda_{(F)}$ is also an invariant subspace of $M$.}:
\begin{eqnarray}
\lambda_{(F)\phantom{E_{2}}i}^{E_{1}E_{2}}\lambda_{(F)E_{2}E_{1}j}+c.c.&=&2m_{E_{4}E_{1}}y_{\phantom{E_{1}}E_{2}i}^{\star E_{1}}m^{E_{2}E_{3}}y_{\phantom{\star}E_{3}\phantom{K}j}^{\star\phantom{E_{2}}E_{4}}+2y_{\phantom{E_{1}}E_{2}i}^{\star E_{1}}m^{E_{2}E_{3}}m_{\phantom{\star}E_{3}E_{4}}^{\star}y_{\phantom{K}E_{1}j}^{E_{4}}+c.c.\nonumber\\
&=&2m_{E_{4}E_{1}}y_{\phantom{E_{1}}E_{2}i}^{\star E_{1}}m^{E_{2}E_{3}}y_{\phantom{\star}E_{3}\phantom{K}j}^{\star\phantom{E_{2}}E_{4}}+2y_{\phantom{E_{1}}E_{2}i}^{\star E_{1}}[m_{(F)}^{2}]_{\phantom{E_{2}}E_{4}}^{E_{2}}y_{\phantom{K}E_{1}j}^{E_{4}}+c.c.\nonumber\\
&=&2m_{E_{4}E_{1}}y_{\phantom{E_{1}}E_{2}i}^{\star E_{1}}m^{E_{2}E_{3}}y_{\phantom{\star}E_{3}\phantom{K}j}^{\star\phantom{E_{2}}E_{4}}+2y_{\phantom{E_{1}}E_{2}i}^{\star E_{1}}\epsilon\delta_{E_{4}}^{E_{2}}y_{\phantom{K}E_{1}j}^{E_{4}}+c.c.\nonumber\\
&=&2y_{\phantom{E_{1}}E_{2}i}^{\star E_{1}}m^{E_{2}E_{3}}y_{\phantom{\star}E_{3}\phantom{K}j}^{\star\phantom{E_{2}}E_{4}}m_{E_{4}E_{1}}+c.c.+O(\epsilon)\nonumber
\end{eqnarray}
From which we finally obtain
\begin{equation}\label{Eq:gammas_to_ys}
\Re\left[\lambda_{(F)\phantom{E_{2}}i}^{E_{1}E_{2}}\lambda_{(F)E_{2}E_{1}j}\right]=2\Re\left[y_{\phantom{E_{1}}E_{2}i}^{\star E_{1}}m^{E_{2}E_{3}}y_{\phantom{\star}E_{3}\phantom{K}j}^{\star\phantom{E_{2}}E_{4}}m_{E_{4}E_{1}}\right]+O(\epsilon)\; .
\end{equation}

\section{Top quark and gauge contributions in the SM}
\label{app:SMcontribs}
In this section we show how to obtain the top-quark couplings to the Higgs boson in the notation set in Sect.~\ref{sec:DefsNotation} in the SM. The Yukawa coupling between the Higgs doublet and the top Quark is given by (using Weyl fermion notation)
\begin{eqnarray}
-\mathcal{L}_{{\rm Yukawa,top}}&=&\, y_tt_R\left(H^c\right)^\dagger T_L+c.c. \label{eq:TopHcoupling}
\\&=&\frac{y_t}{2\sqrt{2}}\left[v t_Rt_L+t_Rt_L\left(h+iG_0\right)-t_Rb_L\left(G_1+iG_2\right)\right]+c.c.\nonumber
\end{eqnarray}
where we have used
\begin{equation}
T_L\equiv \binom{t_L}{b_L}\;,\; H^c\equiv i\sigma_2H^*=\frac{1}{\sqrt{2}}\binom{v+h-i G_0}{-G_+}=\frac{1}{\sqrt{2}}\binom{v+h-i G_0}{-G_1+iG_2}\; .
\end{equation}
Here $t_L,b_L$ and $t_R$ are the left handed Weyl fermions which give, respectively, the left handed part of the top and bottom quarks, and the right handed part of the top quark. Since each fermion is a triplet of colour there is an extra colour contraction between each fermion/anti-fermion pair.
If we organise the three sets of left handed Weyl fermions in a vector 
\begin{equation}
\psi_I\rightarrow(\psi_1,\ldots,\psi_9)=(t_{L}^1,t_{L}^2,t_{L}^3,t_R^1,t_R^2,t_R^3,b_L^1,b_L^2,b_L^3)
\end{equation}
 and the real scalars in a another vector 
\begin{equation}
R_i\rightarrow (h,G_0,G_1,G_2)\; ,
\end{equation}
and note that in the SM this decomposition is already in the mass eigenbasis, we can read $m_{IJ}$ and $y_{IJk}$ directly from Eq.~\eqref{eq:TopHcoupling} with the definitions in Eqs.~\eqref{Eq:lambda_basis}. We can also obtain $\lambda_{(F)IJk}$ and $\lambda_{(F)IJkm}$ using, Eqs.~\eqref{Eq:Lambda-couplings-derivatives}.

The one-loop corrections from the $y_t$ couplings to the pole equations in the SM that are used in the text are written in terms of the following vector, which depends on the momentum-squared scale $s$:
\begin{equation}
S_{\alpha}^{(t)}(s)\rightarrow3y_{t}^{2}\left(\begin{array}{c}
\left(4m_{t}^{2}-s\right)B_{s}\left(m_{t}^{2},m_{t}^{2}\right)-2A\left(m_{t}^{2}\right)\\
-m_{t}^{2}\Delta B_{\bar{F}\bar{F}}\left(m_{t}^{2},m_{t}^{2}\right)-\Delta B_{FF}\left(m_{t}^{2},m_{t}^{2}\right)-6y_{t}^{2}A\left(m_{t}^{2}\right)\\
\Delta B_{FF}\left(0,m_{t}^{2}\right)-6y_{t}^{2}A\left(m_{t}^{2}\right)\\
-\Delta B_{FF}\left(0,m_{t}^{2}\right)-6y_{t}^{2}A\left(m_{t}^{2}\right)
\end{array}\right)
\end{equation}

The scalar-gauge couplings used in the main text result from the following term in the SM Lagrangian
\begin{eqnarray}
-\mathcal{L}_{{\rm V,term}}&=&\, \frac{1}{4}H^\dagger\left(g\sigma^{i}A_{i}^\mu+g'B^\mu\right)\left(gA_{i}^\mu\sigma^i+g'B^\mu\right)H\; ,
\end{eqnarray}
where $g,g'$ are respectively the SU(2) and U(1) couplings in the SM and $A_{i}^\mu,B^\mu$ are the corresponding gauge fields. From this we can then read off the scalar-gauge couplings. Then, we obtain
\begin{eqnarray}
  S_{1}^{(g,1)}(s)&=& \frac{6m_{W}^{4}}{v^{2}}\left(3\logbar m_{W}^{2}+1\right)+\frac{3m_{Z}^{4}}{v^{2}}\left(3\logbar m_{Z}^{2}+1\right) \nonumber\\
  &&+\frac{2m_{W}^{2}}{v^{2}}\left[\Delta B_{SV}\left(0,m_{W}^{2}\right)+2m_{W}^{2}\Delta B_{VV}\left(m_{W}^{2},m_{W}^{2}\right)\right]+ \\
  &&+\frac{m_{Z}^{2}}{v^{2}}\left[\Delta B_{SV}\left(0,m_{Z}^{2}\right)+2m_{Z}^{2}\Delta B_{VV}\left(m_{Z}^{2},m_{Z}^{2}\right)\right]\nonumber \\
  S_{2}^{(g,1)}(s)&=& \frac{2m_{W}^{4}}{v^{2}}\left(3\logbar m_{W}^{2}-1\right)+\frac{m_{Z}^{4}}{v^{2}}\left(3\logbar m_{Z}^{2}-1\right) \nonumber\\
  &&+\frac{2m_{W}^{2}}{v^{2}}\Delta B_{SV}\left(0,m_{W}^{2}\right)+\frac{m_{Z}^{2}}{v^{2}}\Delta B_{SV}\left(0,m_{Z}^{2}\right) \\
  S_{3}^{(g,1)}(s)=S_{4}^{(g,1)}(s)&=&  \frac{2m_{W}^{4}}{v^{2}}\left(3\logbar m_{W}^{2}-1\right)+\frac{m_{Z}^{4}}{v^{2}}\left(3\logbar m_{Z}^{2}-1\right)\nonumber\\
  &&+\frac{m_{W}^{2}}{v^{2}}\left[4\left(1-\frac{m_{W}^{2}}{m_{Z}^{2}}\right)\Delta B_{SV}\left(0,0\right)+\Delta B_{SV}\left(0,m_{W}^{2}\right)+\right.\nonumber\\
    &&+4m_{W}^{2}\left(1-\frac{m_{W}^{2}}{m_{Z}^{2}}\right)\Delta B_{VV}\left(m_{W}^{2},0\right)+\nonumber\\
    &&\left.+4m_{Z}^{2}\left(1-\frac{m_{W}^{2}}{m_{Z}^{2}}\right)^{2}\Delta B_{VV}\left(m_{W}^{2},m_{Z}^{2}\right)\right]+\nonumber\\
  &&+\frac{m_{Z}^{2}}{v^{2}}\left(1-\frac{2m_{W}^{2}}{m_{Z}^{2}}\right)^{2}\Delta B_{SV}\left(0,m_{Z}^{2}\right)
\end{eqnarray}
and

\begin{equation}
  S_{\alpha}^{(g,2)}(m_h^2,s)\rightarrow \left(\begin{array}{c}
0\\
0\\
\frac{m_{W}^{2}}{v^{2}}\Delta B_{SV}\left(m_{h}^{2},m_{W}^{2}\right)\\
\frac{m_{W}^{2}}{v^{2}}\Delta B_{SV}\left(m_{h}^{2},m_{W}^{2}\right)
\end{array}\right)
\end{equation}

\bibliographystyle{JHEP}       
\bibliography{references}   % name your BibTeX data base

\end{document}